\documentclass[11pt,a4paper]{article}
\pdfoutput=1
\usepackage{jheppub}
\usepackage[sort&compress]{natbib}
\usepackage{url}
\usepackage{hyperref}
\usepackage{amsfonts}
\usepackage{epsfig}\usepackage{dcolumn}
\usepackage{bm}
\usepackage{slashed}
\graphicspath{{./figuras/}}
\usepackage{subfigure}
\usepackage{color}

\usepackage{listings}
\usepackage{color}
 
\definecolor{codegreen}{rgb}{0,0.6,0}
\definecolor{codegray}{rgb}{0.5,0.5,0.5}
\definecolor{codepurple}{rgb}{0.58,0,0.82}
\definecolor{backcolour}{rgb}{0.95,0.95,0.92}
 
\lstdefinestyle{mystyle}{
    backgroundcolor=\color{backcolour},   
    commentstyle=\color{blue},
    keywordstyle=\color{codegreen},
    numberstyle=\tiny\color{codegray},
    stringstyle=\color{codepurple},
    basicstyle=\ttfamily\small,
    breakatwhitespace=false,         
    breaklines=true,                 
    captionpos=b,                    
    keepspaces=true,                 
    numbers=left,                    
    numbersep=5pt,                  
    showspaces=false,                
    showstringspaces=false,
    showtabs=false,                  
    tabsize=2
}
 
\newcommand{\bbtautau}{b\bar{b} \tau^{+} \tau^{-}}
\newcommand{\bbWW}{b\bar{b} W^{+} W^{-}}
\newcommand{\bbbb}{b\bar{b} b\bar{b}}

\newcommand{\bbaa}{b\bar{b}\gamma\gamma}
\newcommand{\bbjj}{b\bar{b}jj}
\newcommand{\jjaa}{jj\gamma\gamma}
\newcommand{\bbaj}{b\bar{b}\gamma j}
\newcommand{\ccaj}{c\bar{c}\gamma j}
\newcommand{\ccaa}{c\bar{c}\gamma\gamma}
\newcommand{\tth}{t\bar{t}h}
\newcommand{\zh}{Zh}
\newcommand{\bbh}{b\bar{b}h}
\newcommand{\HyperOpt}{\texttt{Hyperopt }}
\newcommand{\xgb}{\texttt{XGBoost }}
\newcommand{\Delphes}{\texttt{Delphes }}

\title{ {\color{blue} \mbox{Can We Discover Double Higgs Production at the LHC?}}}

\author{Alexandre Alves$^a$,}
\author{Tathagata Ghosh$^{b}$,}
\author{Kuver Sinha$^{c,d}$}

\affiliation{$^a$Departamento de F\'isica, Universidade Federal de S\~ao Paulo, Diadema-SP, 09972-270, Brazil}
\affiliation{$^b$Department of Physics and Oklahoma Center for High Energy Physics, \\ Oklahoma State University, Stillwater, OK 74078-3072, USA}
\affiliation{$^c$Department of Physics and Astronomy, University of Oklahoma, Norman, OK 73019, USA}
\affiliation{$^d$Department of Physics and Astronomy, University of Utah, Salt Lake City, UT 84112, USA}

\emailAdd{aalves@unifesp.br}
\emailAdd{tghosh@okstate.edu}
\emailAdd{kuver.sinha@ou.edu}

\abstract{We explore double Higgs production via gluon fusion in the $b\bar{b} \gamma \gamma $ channel at the high-luminosity LHC using machine learning tools. We first propose a Bayesian optimization approach to select cuts on kinematic variables, obtaining a $30-50$ \% increase in the significance compared to current results in the literature. We show that this improvement persists once systematic uncertainties are taken into account. We next use boosted decision trees (BDT) to further discriminate signal and background events. Our analysis shows that a joint optimization of kinematic cuts and BDT hyperparameters results in an appreciable improvement in the significance. Finally, we perform a multivariate analysis of the output scores of the BDT. We find that assuming a very low level of systematics, the techniques proposed here will be able to confirm the production of a pair of Standard Model Higgs bosons at 5$\sigma$ level with 3 ab$^{-1}$ of data. Assuming a more realistic projection of the level of systematics, around 10\%, the optimization of cuts to train BDTs combined with a multivariate analysis delivers a respectable significance of 4.6$\sigma$. Even assuming large systematics of 20\%, our analysis predicts a 3.6$\sigma$ significance, which represents at least strong evidence in favor of double Higgs production. We carefully incorporate  background contributions coming from light flavor jets or $c$-jets being misidentified as $b$-jets and jets being misidentified as photons in our analysis.}
    
\begin{document} 

\begin{flushright}
OSU-HEP-17-01
\end{flushright}

\maketitle
\flushbottom
    
\section{Introduction}

Measuring possible deviations of the triple Higgs coupling $\lambda_3$ from its predicted Standard Model (SM) value is a key goal of future colliders. This has implications for a whole range of new physics scenarios, such as supersymmetry and other extensions of the SM with two Higgs doublets. The cosmological implications are also profound, since $\lambda_3$ is related to the strength of the electroweak phase transition which is critical for understanding electroweak baryogenesis, for example. 

The triple Higgs coupling can be probed by Higgs pair production processes, which have been extensively studied in the context of the high-luminosity LHC and future hadron colliders. Higgs pair production occurs dominantly via gluon fusion, with other production processes being more than an order of magnitude smaller. Final states that have been studied, in the context of di-Higgs production at the LHC, include $\bbaa$ ~\cite{Baur:2003gp, Baglio:2012np, Huang:2015tdv, Azatov:2015oxa, Barger:2013jfa, ATLAS14, ATLAS17}, $\bbtautau$ \cite{Baur:2003gpa, Dolan:2012rv}, $\bbWW$ \cite{Papaefstathiou:2012qe}, and $\bbbb$ \cite{deLima:2014dta, Wardrope:2014kya, Behr:2015oqq}.

The purpose of this paper is to investigate the prospects of Higgs pair production at the LHC in the $\bbaa$ channel. Our analysis builds on previous studies in two ways: we use tools from the machine learning (ML) literature in our analysis, and we carefully account for background contributions coming from light flavor jets ($j$) or $c$-jets being misidentified as $b$-jets and electrons or jets being misidentified as photons. With regard to the use of ML tools, we note that this is somewhat hostile terrain for theorists. However, the comparative gains in discovery prospects over other methods, which we discuss at length, will hopefully convince the reader that planning for future colliders should exploit state of the art data analysis tools to ensure that projections are reasonable.

For the benefit of the reader, we chart out the steps in our analysis and the main results of each step. We present the details of our signal and background simulation in section \ref{section:simulation}. We provide a brief discussion on previous studies in section \ref{section:compare}.   

In section \ref{section:optimization}, we ask the question: given an event topology and a set of kinematic observables, is there a systematic and computationally feasible method to obtain the most optimal selections that maximize the significance? We show that Bayesian optimization, as described in refs.~\cite{Bergstra1, Bergstra2}, performs better than selections currently proposed in the literature, and is computationally much more tractable than a brute force multivariable scan. We demonstrate our results with the Python algorithm \HyperOpt~\cite{url:hyperopt}. Our main results of this section are presented in figure~\ref{fig:2} and we find that there is a $30-50$ \% increase in the significance metric $S/\sqrt{B}$ compared to current results in the literature. Moreover, this relative improvement persisted after incorporating systematic uncertainties on the background rate, as demonstrated in figure~\ref{fig:3}.

In section \ref{section:BDT}, we build on the Bayesian optimization of kinematic cuts, and show that training a Boosted Decision Tree (BDT) algorithm to better classify signal and backgrounds events, in addition to the procedure of using optimal cuts to select the best volume of the features space for the BDT training, increases the discovery prospects dramatically. For our calculations, we use the \xgb~\cite{TChen} implementation of BDTs for Python. We present our results in three stages. In section~\ref{section:BDTcuts}, we first introduce the kinematic observables used in the BDT analysis, and provide a discussion of the interplay between BDT classifiers and cut selections, without addressing the question of cut optimization. In section~\ref{section:sequential} we sequentially optimize the cuts on the kinematic observables using \HyperOpt, and then optimize the BDT hyperparameters. Finally, in section~\ref{section:joint}, we perform a joint optimization of the kinematic cuts and the BDT hyperparameters. 

Our results from this stage of the analysis are summarized in table~\ref{table:4} and figure~\ref{fig:9}. We find that the use of BDT enhances the significance irrespective of the kinematic cuts used. The largest enhancement, however, occurs with cuts optimized using \HyperOpt, and we reach a significance of $3.88$ for 3000 fb$^{-1}$ of data.

In section \ref{section:mva}, we focus on the statistical side of the analysis by estimating the log-likelihood ratio statistics from the output scores of the BDTs provided by \xgb, following~\cite{Cranmer:2016swd}. 

The final results of our paper are presented in table~\ref{table:6}. We find that assuming a very low level of systematics, the techniques proposed here will be able to confirm the production of a pair of SM Higgs bosons at 5$\sigma$ level. Assuming a more realistic projection of the level of systematics, around 10\%, the optimization of cuts to train BDTs combined with a multivariate analysis delivers a respectable significance of 4.6$\sigma$. This is the largest significance achieved so far in the $\bbaa$ channel with realistic assumptions concerning backgrounds and systematic uncertainties at the 14 TeV LHC. Even assuming large systematics of 20\%, our analysis predicts a 3.6$\sigma$ significance, which represents at least strong evidence in favor of double SM Higgs production.

We pause to make a few comments about signal and background event rate estimation before proceeding with our analysis. There has been considerable disagreement about this in the literature, with some of the older studies giving optimistic results due to an underestimation of background. We discuss these issues in section~\ref{section:compare}, where we compare and summarize previous studies. Throughout this work, we will take the background and signal event rates of Azatov \emph{et. al.}, ref.~\cite{Azatov:2015oxa}, which we consider robust, as a reference point. However, we are also careful to incorporate the backgrounds $c\bar{c}\gamma\gamma$, $b\bar{b}\gamma j$ and $c\bar{c}\gamma j$, whose importance has been highlighted by ATLAS ref.~\cite{ATLAS14}. 

 In appendix~\ref{app:1} we briefly comment about the metrics used to compute the statistical significances, and in appendix~\ref{app:2} we show a Python snippet of a simple code to implement the selection cuts optimization based on \HyperOpt. 

\section{Details of $pp\to b\bar{b}\gamma\gamma$ simulations}
\label{section:simulation}

The details of the signal and background simulation will be presented in this section.

Instead of re-evaluating all cross sections for the process of interest, the strategy we will pursue in this work is to assume the production rates presented in ref.~\cite{Azatov:2015oxa}. In our opinion, the calculations performed by Azatov \emph{et. al.} are reliable enough to be used as a starting point, especially given that we are interested in a close comparison of our results with those previously obtained in the literature. We will use events generated only as a means of estimating the kinematic distributions germane to the cut-and-count analysis and to train our ML algorithms. We do, however, take into account three additional sub-dominant backgrounds beyond those of ref.~\cite{Azatov:2015oxa}. 


\subsection{Higgs pair production}


For the simulation of the signal and backgrounds events, we use {\tt MadGraph5\_aMC@NLO\_v2.3.3} \cite{MG5} with the CTEQ5L~\cite{Lai:1999wy} and CTEQ6L~\cite{Nadolsky:2008zw} parton distribution functions, respectively. At the leading order (LO), there are two one-loop diagrams that contribute to the process $pp\to hh$~\cite{Glover:1987nx, Dicus:1987ic, Plehn:1996wb} and they interfere destructively. While the triangle diagram is sensitive to the Higgs trilinear coupling, $\lambda_3$, the box diagram is not. The simulation of our signal includes the effect of both these diagrams. However, over the last 30 years significant improvement on the theoretical calculation of this process to higher orders~\cite{Dawson:1998py, Kniehl:1995tn, deFlorian:2013uza, deFlorian:2013jea, Grigo:2014jma, deFlorian:2015moa, Shao:2013bz} has taken place. 

In ref.~\cite{Azatov:2015oxa}, the signal cross section at the 14 TeV LHC was calculated at LO with {\tt MadGraph5\_aMC@NLO\_v2.1.1} and then multiplied by the partial NNLO K-factor of 2.27~\cite{deFlorian:2013uza}, calculated in the large quark mass limit. The resulting production cross section is 36.8 fb. The combined branching ratio of the $\bbaa$ channel is small, only 0.264\%. The number of signal events after 3000 fb$^{-1}$, before cuts and efficiencies, is around 290.

Effects of the finite top quark mass to the NLO QCD cross section of Higgs pair production has been taken into account in refs.~\cite{Borowka:2016ehy,Borowka:2016ypz}. The full mass dependence diminishes the NLO prediction by 14\% compared to the large top quark mass approximation, however approximated NNLO effects increase the NLO predictions by $\sim 20$\% according to ref.~\cite{deFlorian:2016uhr}, therefore, the K-factor adopted by Azatov \emph{et. al.} constitutes a fair approximation to the total rate. 

Hard jet radiation and finite top quark mass effects are also expected to change the shape of distributions involving the four-momenta of the reconstructed Higgs bosons at higher orders as shown in refs.~\cite{Borowka:2016ehy,Borowka:2016ypz,deFlorian:2016uhr,Heinrich:2017kxx}. 

In order to obtain the distributions of the kinematic variables of interest, we pass our simulated events to {\tt PYTHIA\_v6.4~\cite{pythia} for showering and hadronization.} Finally, these events are passed to {\tt DELPHES\_v3.3}~\cite{delphes} for detector simulation. For the signal, the Higgs bosons are decayed into bottom quarks and photons with the \texttt{MadSpin} module of \texttt{MadGraph5}. In contrast, for the relevant backgrounds which contain a Higgs in the final state, the Higgs boson has been decayed within {\tt PYTHIA}. Photon isolation criteria and jet clustering are similar of those of Azatov \emph{et. al.} who found that their results do not differ much from other works with somewhat different criteria.


Both signals and backgrounds were required to pass the following minimal selection criteria
\begin{eqnarray}
& & p_T(j) > 20\;\hbox{GeV},\; p_T(\gamma) > 20\;\hbox{GeV},\; |\eta(j)|<2.5,\; |\eta(\gamma)|<2.5\\
& & 100\;\hbox{GeV} < |M_{jj}| < 150\;\hbox{GeV},\; 100\;\hbox{GeV} < |M_{\gamma\gamma}| < 150\;\hbox{GeV}\; .
\label{cuts:basic} 
\end{eqnarray}
In the next section we comment about the backgrounds and give further details of the computations.

It is important to stress that a better estimation of production rates and invariant mass distributions would mainly require including the effects of the finite top quark mass and higher order corrections. That, however, is beyond the scope of this work. 


\subsection{Backgrounds}

We have evaluated the backgrounds to $(h \rightarrow b\bar{b}) +  (h \rightarrow \gamma\gamma)$ signal from multiple SM processes:
\begin{enumerate}
\item $\bbaa$;
\item $\zh$, $Z\to b\bar{b}$ and $h\to\gamma\gamma$;
\item $\bbh$, $h\to\gamma\gamma$;
\item $\tth \rightarrow b\bar{b}+\gamma\gamma+X$;
\item $\jjaa$, where the light-jets $jj$ are mistaken for a $b$-jet pair in the detector;
\item $\bbjj$,  where the light-jets $jj$ are mistaken for a photon pair in the detector;
\item $\ccaa$, where a $c$-jet is mistagged as a $b$-jet;
\item $\bbaj$, one light-jet is mistaken for a photon;
\item $\ccaj$, the $c$-jets are mistagged as bottom jets and the light-jet as a photon.
\end{enumerate} 

The cross section normalizations for the backgrounds from 1 to 5 are taken from ref.~\cite{Azatov:2015oxa}. In that work, the continuum $\bbaa$ is computed at LO with one extra jet radiation and a K-factor of 2 is estimated for the NLO QCD corrections.This large K-factor for the dominant background has been neglected in many previous studies in this channel. The backgrounds $\zh$ and $\bbh$ were also evaluated with one extra jet radiation to estimate the higher-order QCD corrections. The $\tth$ K-factor was taken from~\cite{HWG} and it is small. The signal and backgrounds estimates of Azatov \emph{et. al.} are found to agree reasonably well of the Snowmass group report of ref.~\cite{Dawson:2013bba}.

Our backgrounds events (1--4) are also generated with 1 extra parton radiation in order to better simulate the kinematic distributions. MLM scheme~\cite{MLM} of jet-parton matching has been utilized to avoid double counting. The extra hard jet was included in the $\bbaa$ background once it is the dominant one. The reason for including the extra QCD radiation in the ressonant backgrounds $\tth$, $\zh$ and $\bbh$ is that the Higgs boson recoils against the extra hard jets which is important to obtain the $M_{\bbaa}$ invariant mass distribution. Unfortunately, it is computationally too expensive to simulate the signals in the same way, and beyond our means.

The $\tth$ background is simulated in the inclusive way. Events with hard charged leptons are easily classified as backgrounds events however and efficiently discarded as we are going to see. 

Background processes with light jets are important when a jet radiates a hard photon which is mistaken for an isolated photon in the detector. This is the case of the backgrounds $\bbjj$, $\bbaj$ and $\ccaj$. All the backgrounds from 5 to 9 in the above list were simulated with {\tt MadGraph5\_aMC@NLO\_v2.3.3} at LO  and multiplied by the NLO QCD K-factors presented in ref.~\cite{Alwall:2014hca}. 

Following previous studies~\cite{Baur:2003gp, Baglio:2012np, Huang:2015tdv, Azatov:2015oxa, Barger:2013jfa, ATLAS14}, we adopt the probability of $1.2\times 10^{-4}$ for a light-jet to be mistagged as a photon. However, in the presence of pile-up events this value might be an underestimate~\cite{ATLAS17}. Nevertheless, the $\bbjj$ background was found to be negligible after imposing cuts and mistagging factors.

Finally, for $\ccaa$ backgrounds where a $c$-jet is mistagged as a $b$-jet, the $b$ and $c$-tagging, and also the light-jet mistagging are parametrized according to the jet's transverse momentum and rapidity as implemented in \Delphes, specifically as the the default simulation of the CMS detector. The \Delphes parametrization assumes that a 70\% $b$-tagging efficiency is reached for $p_T>100$ GeV at the cost of a 20(5)\% mistagging factor for $c$($j$)-jets. These sub-dominant backgrounds $\bbaj$, $\ccaa$ and $\ccaj$ were not taken into account in the majority of the previous studies we are considering in this work for comparisons, except for \cite{Huang:2015tdv,ATLAS14,ATLAS17}. All the uncertainties in the backgrounds rates are taken into account in this work as systematic uncertainties in the calculation of the signal significances.

The numbers of backgrounds events after imposing the basic cuts of eq.~(\ref{cuts:basic}) for 3 ab$^{-1}$ of integrated luminosity is shown in table~\ref{table:nev}.
\begin{table}[t]
\centering
\begin{tabular}{c|c|c|c|c|c|c|c|c|c}
\hline\hline
signal & $\bbaa$ & $\ccaa$ & $\jjaa$ & $\bbaj$ & $\tth$ & $\ccaj$ &  $\bbh$ & $\zh$ & total backgrounds \\ 
\hline
42.6   & 1594.5  & 447.7   &  160.3 &   137 & 101.1  & 38.2  &  2.4    & 1.8 & 2483 \\
\hline\hline
\end{tabular}
\caption{The number of signal and the various types of backgrounds considered in this work after imposing the basic cuts of eq.~(\ref{cuts:basic}) for 3 ab$^{-1}$ of data. We found $\bbjj$ negligible after cuts and estimating the probability of the jet pair fakes a photon pair.}
\label{table:nev}
\end{table}
%


In the next section, we will investigate a method to optimize the cut-and-count analysis, instead of manual tuning of cut thresholds as is commonly done. This requires us to plant ourselves on a set of baseline results and cut strategies, but also to adopt the signal and background normalizations of this baseline work. We chose to adopt the results, cuts and normalization of ref.~\cite{Azatov:2015oxa} as our baseline due their careful treatment of signals and backgrounds concerning QCD higher order effects. As we will show, this work also presents the best cut strategy when compared to other theoretical and experimental works. On the other hand, we go beyond that work by including the sub-dominant backgrounds $\bbaj$, $\ccaa$ and $\ccaj$. Our simulations for these backgrounds agree reasonably well with those from \cite{Huang:2015tdv,ATLAS14,ATLAS17}.

\section{Comparison and Summary of Previous Studies}
\label{section:compare}

In this section, we present a summary of previous studies of double Higgs production at the LHC, taking refs.~\cite{Baur:2003gp, Baglio:2012np, Huang:2015tdv, Azatov:2015oxa, Barger:2013jfa, ATLAS14} as representatives. Our main goal here is to show that despite the varying levels of rigor in terms of calculating signal and background event rates, and the differences in selection strategies, the cut and count analyses employed in these disparate studies yield similar significances.

The process $pp\to hh\to \bbaa$, with final states containing two $b$-jets and two hard photons, presents many features which make it possible to employ a large variety of kinematic variables and selection strategies. We describe the most pertinent ones below:


%
\begin{enumerate}
\item transverse momentum of $b$-jets and photons: $p_T(b)$ and $p_T(\gamma)$
\item $b\bar{b}$ and $\gamma\gamma$ invariant masses: $M_{bb}$ and $M_{\gamma\gamma}$, where signal events exhibit resonance peaks at $m_h$
\item transverse momentum of $b\bar{b}$ and $\gamma\gamma$: $p_T(bb)$ and $p_T(\gamma\gamma)$
\item invariant mass of two $b$-jets and two photons: $M_{bb\gamma\gamma}$
\item distance between pairs of $b$-jets and photons: $\Delta R(bb)$, $\Delta R(\gamma\gamma)$ and $\Delta R(b\gamma)$, where
$\Delta R=\sqrt{(\Delta\eta)^2+(\Delta\phi)^2}$ in the pseudo-rapidity and azimuthal angle plane $(\eta,\phi)$
\item the fraction $E_T/M_{\gamma\gamma}$ for the two hardest photons in the event; these are variables used in experimental searches as in ref.~\cite{Aad:2014yja, CMS}
\end{enumerate}

\begin{figure}[!t]
\centering
\subfigure[\label{fig01}]{\includegraphics[scale=0.35]{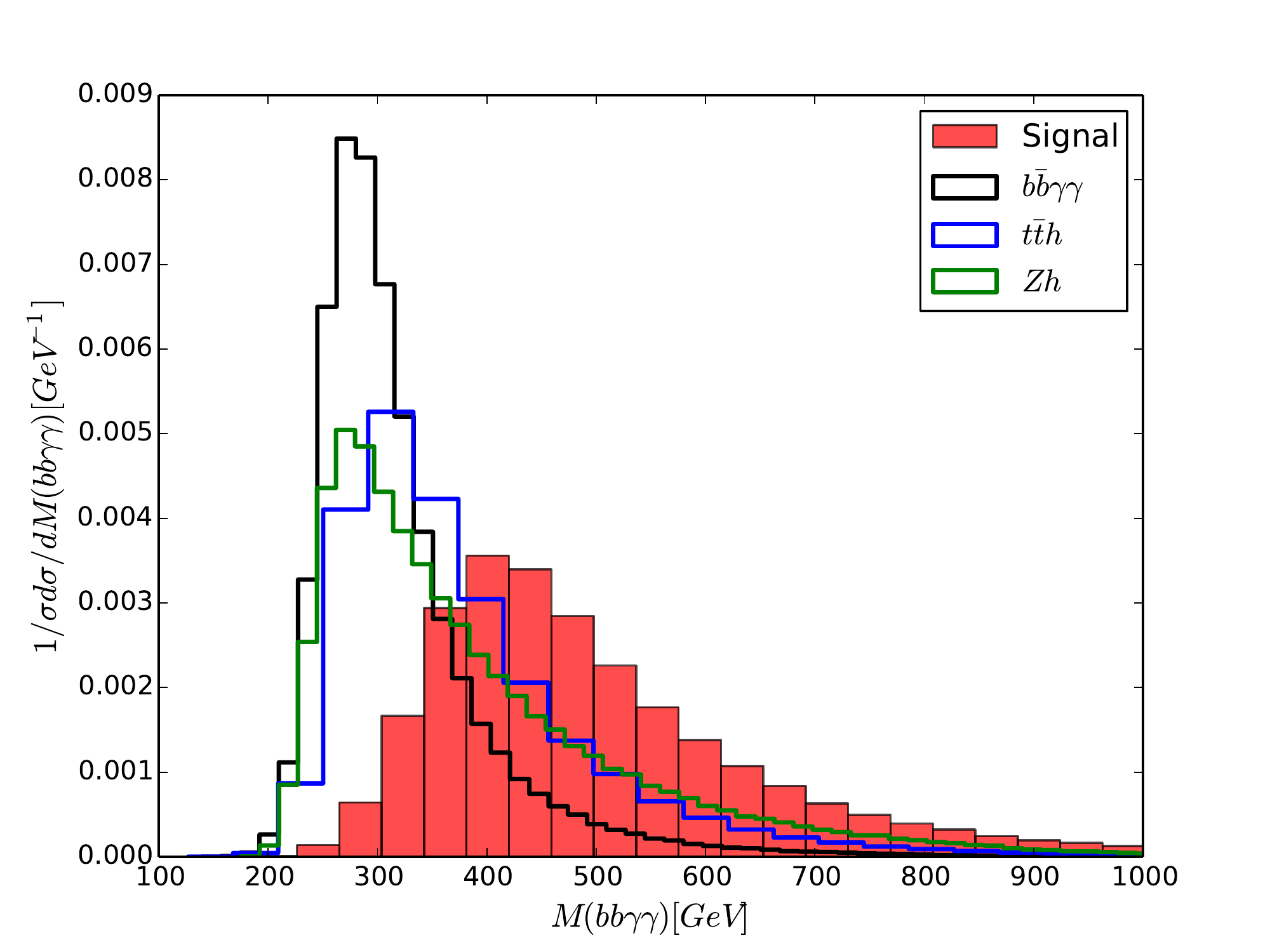}}
\subfigure[\label{fig02}]{\includegraphics[scale=0.35]{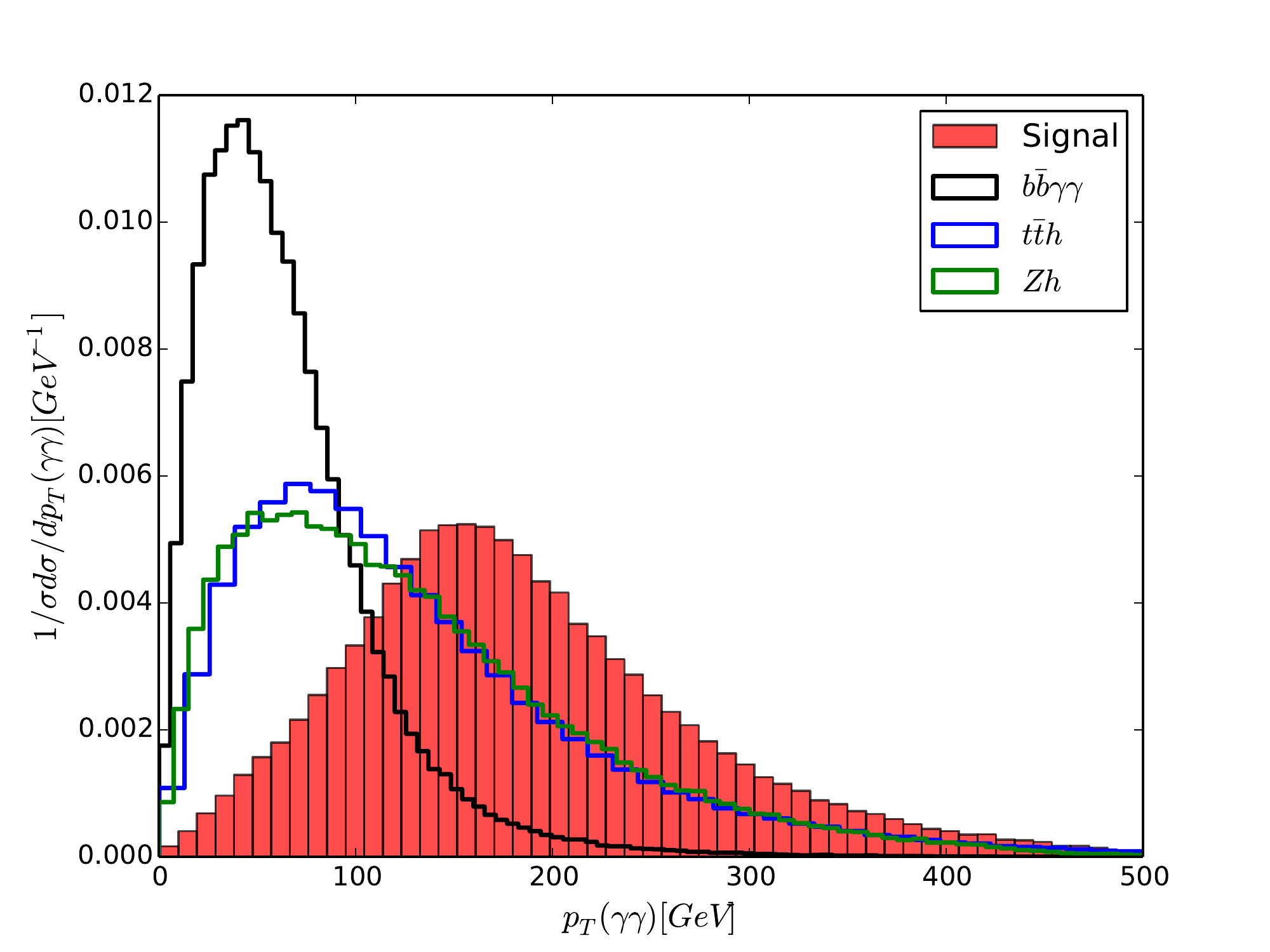}}\\
\subfigure[\label{fig03}]{\includegraphics[scale=0.35]{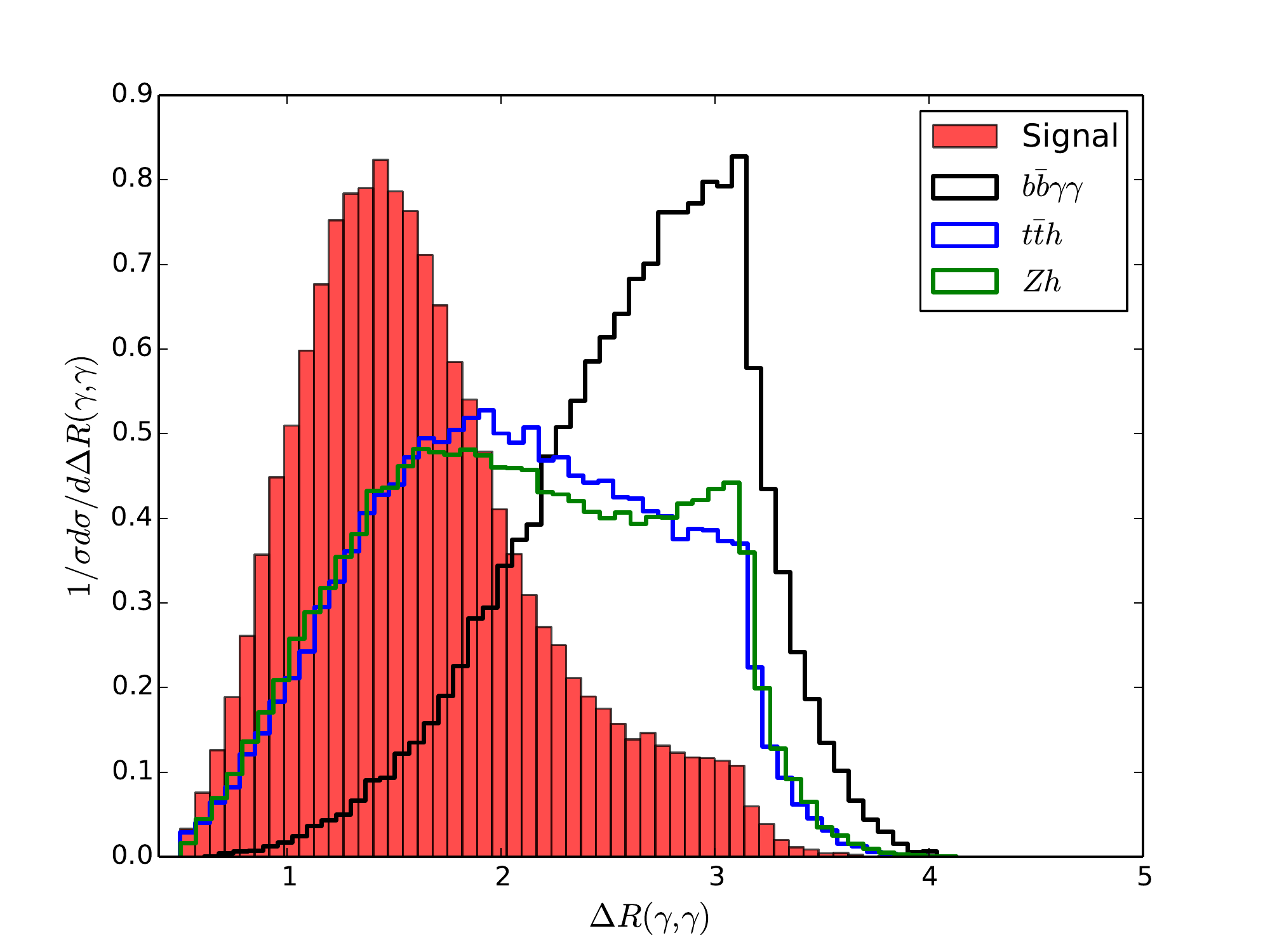}}
\subfigure[\label{fig04}]{\includegraphics[scale=0.35]{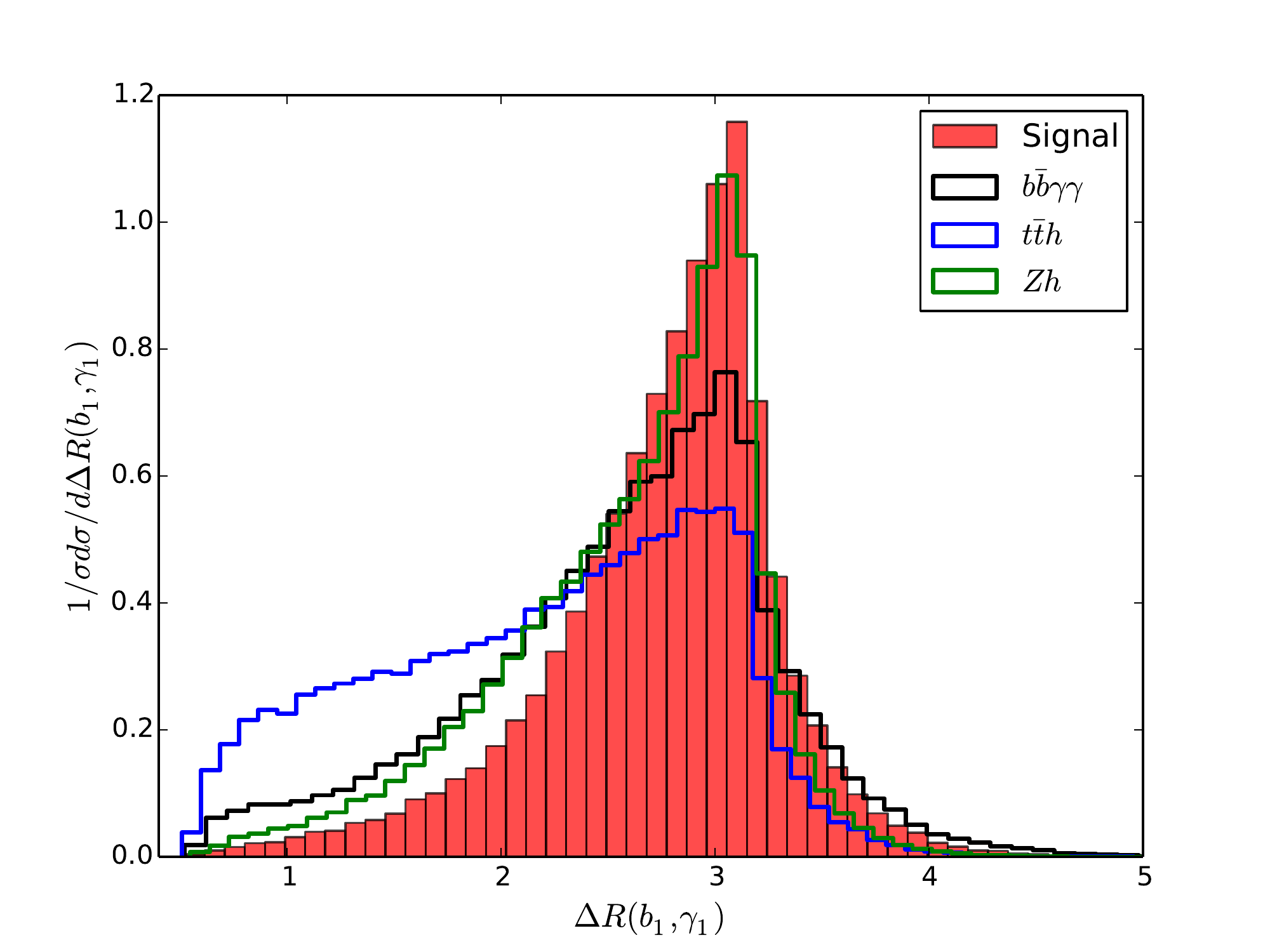}}
\caption{Kinematic distributions of the signal (shaded red), and the backgrounds $b\bar{b}\gamma\gamma$ (black), $t\bar{t}h$ (blue) and $Zh$ (green) are displayed. In (a), we show the invariant mass of two $b$-jets and two photons. In (b), we show the transverse momentum of a pair of photons. In (c) and (d), we show the distance $\Delta R$ between a pair of photons, and between the hardest photon and the hardest $b$-jet, respectively.}
\label{fig:1}
\end{figure}

Some of these kinematic distributions have been presented in figure~\ref{fig:1} for the signal, and continuum $b\bar{b}\gamma\gamma$, $t\bar{t}h$ and $Zh$ backgrounds.  In panel (a), we show the invariant mass of two $b$-jets and two photons. In panel (b), we show the transverse momentum of a pair of photons. In panels (c) and (d), we show the distance $\Delta R$ between a pair of photons, and between the hardest photon and the hardest $b$-jet, respectively.

In table~\ref{table:1}, we display the analyses performed by the representative theory groups, along with the ATLAS study~\cite{ATLAS14}, which is shown in the last row.  The first column gives the relevant reference, while the second column gives the kinematic variables and selections that were applied in the corresponding paper. The different groups made very different signal and background estimates, and we refer to ref.~\cite{Azatov:2015oxa} for a detailed discussion of these differences. For the significance calculations shown in the final column, we take all signal and background cross sections to be normalized to the values obtained by ref.~\cite{Azatov:2015oxa}, which, in our opinion, is the most robust theory study. However, we also take into account the backgrounds $c\bar{c}\gamma\gamma$, $b\bar{b}\gamma j$ and $c\bar{c}\gamma j$ which were not taken into account in ref.~\cite{Azatov:2015oxa}. 

The final column of table~\ref{table:1} thus shows the performance that each group would have had with its selection strategies, if all cross-sections had been normalized by the ones of ref.~\cite{Azatov:2015oxa} and if $c\bar{c}\gamma\gamma$, $b\bar{b}\gamma j$ and $c\bar{c}\gamma j$ backgrounds had been taken into account. The statistical significance for each study is calculated with the naive metrics of eq.~(\ref{sig:naive}), for 3 ab$^{-1}$  of data with no systematic uncertainties. The numbers inside parenthesis denote the $S/B$ ratio of each study.

Our main message from table~\ref{table:1} is that the different search strategies employed by the groups yield similar significances, once signal and background cross sections are normalized to the proper value. In other words, the selections and cut and count analysis of any particular group does not radically outperform that of any other.

\begin{table}[t]
\centering
\begin{tabular}{c|c|c}
\hline
Reference & Kinematic cuts & AMS($\sigma$) ($S/B$)\\ 
\hline \hline
& $p_{T_{\gamma(b)}}>20(45)$ GeV, $|\eta_{b,\gamma}|<2.5$ &  \\
(A) \cite{Baur:2003gp} & $|M_{bb}-m_h|<20$ GeV, $|M_{\gamma\gamma}-m_h|<2.3$ GeV &  1.54(0.30)\\
 & $\Delta R_{b\gamma}>1.0$, $\Delta R(\gamma\gamma)<2.0$ & \\
\hline
 & $p_{T_{b,\gamma}}>50$ GeV, $|\eta_{b,\gamma}|<2.5$,  $\Delta R_{b\gamma}>0.4$, $\Delta R(bb)<2.5$ &  \\ 
(B) \cite{Baglio:2012np} & $110<M_{bb}<135$ GeV, $|M_{\gamma\gamma}-m_h|<5$ GeV, $M_{bb\gamma\gamma}>350$ GeV & 1.33(0.39) \\
 & $|\eta_H|<2$, $P_{T_H} > 100$ GeV  &  \\
\hline
 & $p_{T_{b,\gamma}}>30$ GeV, $|\eta_{b,\gamma}|<2.5$ & \\
(C) \cite{Huang:2015tdv} & $|M_{bb}-m_h|<12.5$ GeV, $|M_{\gamma\gamma}-m_h|<5$ GeV & 1.51(0.17) \\
 & $M_{bb\gamma\gamma}>350$ GeV &  \\
\hline
 & $p_{T_{1(2)}}>30(50)$ GeV, $|\eta_{b,\gamma}|<2.4$   &  \\
(D) \cite{Azatov:2015oxa} & $\Delta R_{b\gamma}>1.5$, $\Delta R(bb,\gamma\gamma)<2$  &  1.76(0.27) \\
 & $|M_{bb}-m_h|<20$ GeV, $|M_{\gamma\gamma}-m_h|<5$ GeV & \\
\hline
& $p_{T_{\gamma}}>30(30)$ GeV, $p_{T_{\gamma}}>40(25)$ GeV, $|\eta_{b,\gamma}|<2.4$ & \\
ATLAS \cite{ATLAS14} & $\Delta R_{b\gamma}>0.4$, $\Delta R(bb,\gamma\gamma)<2$, $p_{T_{bb,\gamma\gamma}}>110$ GeV &  1.73(0.28) \\
 & $|M_{bb}-m_h|<25$ GeV, $123<M_{\gamma\gamma}<128$ GeV & \\
\hline\hline
\end{tabular}
\caption{In the first column at left we show the literature references of each cut strategy displayed at the second column. In the last column we compute the signal significance with the number of signal and background events estimated in this work. The number inside parenthesis in the last columns are the signal-to-background ratios. We took the $\ccaa$, $\bbaj$ and $\ccaj$ backgrounds into account but no systematics. The Approximated Mean Significance (AMS) function significance is that of eq.(\ref{sig:ams}).}
\label{table:1}
\end{table}

We now discuss the studies conducted by the different groups in more detail.

The sets (A) and (D), from refs.~\cite{Baur:2003gp} and \cite{Azatov:2015oxa}, displayed in the first and fourth rows of table~\ref{table:1}, respectively, rely on the very distinctive shapes of the $\Delta R_{bb},\; \Delta R_{\gamma\gamma}$ and $\Delta R_{b\gamma}$ distributions to reduce background events. In plot (c) of figure~\ref{fig:1} we show the $\Delta R_{\gamma\gamma}$ distribution for the signal and the main backgrounds. For the signal, photons come from the decay of a heavy particle and are more collimated with diminished distance in the $(\eta,\phi)$ plane. On the other hand, the photons and $b$-jets from the $\bbaa$ continuum originate from QED and QCD radiation, respectively, and are thus less collimated. The $\tth$ and $Zh$ backgrounds resemble the signal as they contain a Higgs boson. The same occurs for the $\Delta R_{bb}$ distribution except for $\tth$ as the $b$-jets come from different top decays. The $\Delta R_{b\gamma}$ distribution between the hardest $b$-jet and photon, shown in the panel (d) of figure~\ref{fig:1}, is more useful to reduce the $\tth$ backgrounds since the bottoms from top decays and the radiated photons from them tend to get more collimated.

 The sets (B) and (C), from refs.~\cite{Baglio:2012np, Huang:2015tdv}, displayed in the second and third rows of table~\ref{table:1}, respectively, take advantage of the fact that the signal events feature a harder spectrum of the $\bbaa$ invariant mass and the transverse momentum of the $b$-jet and photon pair distributions, $p_T(bb)$ and $p_T(\gamma\gamma)$. This is evident from panels (a) and (b) of figure~\ref{fig:1}. We note that these strategies, however, do not reach a higher efficiency compared to those based solely on $\Delta R$ distributions. Moreover, the $S/B$ ratio also does not differ significantly. The set (B) is able to reach almost 0.4, but at the expense of accepting more backgrounds which decreases the significance with no systematics compared to the other analyses. This conclusion may be somewhat modified if systematic uncertainties are incorporated.

The set of cuts from ATLAS combines selections across all the theoretical studies, as can be seen from the last row of table~\ref{table:1}. A signal significance of $\sim 1.73\sigma$, very similar to that of set (D) from ref.~\cite{Azatov:2015oxa}, is obtained. 

It is interesting to compare the signal and background yields obtained in our work to those of the ATLAS paper, ref.~\cite{ATLAS14}. Adopting the cuts of the last row of table~\ref{table:1}, we found 11.8 and 41.9 events for signal and backgrounds, respectively, compared to 8.4 for the signal and 47.1 for backgrounds quoted in ref.~\cite{ATLAS14}. The $S/\sqrt{B}$ significance of the ATLAS paper, with 7.5\% systematics in the background rate quoted in that study, is 1.3$\sigma$, against 1.6$\sigma$ from our results. This cross check gives us confidence in our signal and background estimates and reassure the importance of including systematic uncertainties. The discrepancy between our estimates and those of ATLAS may in part be explained by the fact that we do not discard photons which hit the Barrel/End-Cap transition region, $1.37<|\eta|<1.52$, and are totally inclusive in the number of jets accepted.  The ATLAS study, on the other hand, included events up to 5 jets with $p_T > 25$ GeV. These somewhat looser criteria might explain part of the discrepancy between our estimates.

 More recently, the ATLAS Collaboration updated the prospects for this channel in ref.~\cite{ATLAS17} taking pile-up effects and some other sub-dominant backgrounds, such as $Z(\to b\bar{b})\gamma\gamma$ and $t\bar{t}\gamma$, into account. Pile-up effects were shown to have moderate influence in the discovery prospects, but the backgrounds were found to be somewhat larger than before. The signal significance is estimated to be approximately $1\sigma$ for around 8\% systematics in the background rate with the $S/\sqrt{B}$ metrics. The major discrepancy compared to the previous ATLAS study of ref.~\cite{ATLAS14} and other works is in the number of $\bbaj$ events, which was estimated to be almost as large as $\bbaa$ due to an estimated probability for a jet to fake photons that was four times larger than that  assumed in previous studies. Since we do not take into account the effect of pile-up, we keep comparing our results against those of ref.~\cite{ATLAS14}.

The cut strategy in this new ATLAS study ref.~\cite{ATLAS17} followed  the previous study of ref.~\cite{ATLAS14} closely. The main difference was a softening of the $p_T(b\bar{b},\gamma\gamma)$ cut by vetoing events where this variable is less than 80 GeV. The significance obtained after applying these cuts with our extended backgrounds, assuming no systematics and using the AMS metric, is $1.76\sigma$ and $S/B=0.26$. This is very similar to the results of the last row of table~\ref{table:1}.

Finally, since the subsequent sections will be devoted to applications of ML algorithms to the question of Higgs pair production, we note that in ref.~\cite{Barger:2013jfa}, a likelihood function-type discriminator was built to better discriminate between signal and background events with a large improvement in the signal significance. In that work, however, an underestimation of backgrounds led to a large significance not confirmed in subsequent analyses.

\section{Optimal Selection of Kinematic Cuts}
\label{section:optimization}

In the previous section, we discussed the analysis performed by several theory groups, as well as an ATLAS study. The summary is provided in table~\ref{table:1}, where it is evident that once signal and background cross sections are properly accounted for, the studies are similar in their performances. 

The similarity among the performances of refs.~\cite{Baur:2003gp, Baglio:2012np, Huang:2015tdv, Azatov:2015oxa, ATLAS14} shown in table~\ref{table:1} suggests that the quest for superior performance in cut and count analyses is largely based on previous results and well known variables proposed in the literature. Sometimes, new variables are found to exhibit good discriminative power, such as the ratio $E_T(\gamma)/M_{\gamma\gamma}$ proposed in ref.~\cite{CMS}. Of course, there is a lot of variation in the way different groups design their cuts, the extent to which they experiment with old and new variables, and the methods they employ to estimate the boundary of the chosen kinematic variables


\subsection{Bayesian Optimization of Kinematic Cuts}
\label{section:smbo}
Given an event topology and a set of kinematic observables, is there a way to systematically obtain the most optimal cuts on the kinematic observables, so as to maximize the significance? Our purpose in this section is to probe this question, and we shall see that Bayesian optimization offers a pathway.

A typical cut analysis consists in finding a set of kinematic variables thresholds $\{x_k^{c},k=1,\cdots, n\}$ such that
the number of signal or background events is given by
\begin{equation}
S,B(x_1^c,\cdots,x^c_{n})=L\times\sigma_{S,B}(pp\to X)\times \varepsilon_{eff}\times \prod_{k=1}^{n} H({\cal O}_k(x_k,x_k^c))
\label{nevents:1}
\end{equation}
where $L$ is the integrated luminosity, $\sigma_{S,B}$ is the signal or background production cross section of $X$, $\varepsilon_{eff}$ a factor that accounts for detection efficiencies, and $H$ is the Heaviside step function. The functions ${\cal O}(x,y)$ relate a kinematic variable $x$ and its cut $x^c$ according to one of the following alternatives in this work: $x-x^c$, $x^c-x$, and $|x-M|-x^c$. The goal of our phenomenological analysis is of maximizing a signal significance metric, such as $S/\sqrt{B}$, by retaining the largest possible number of signal events while rejecting the largest amount of background events by finding an optimal set of cuts $\{x_k^{c},k=1,\cdots, n\}$.  

When a ML algorithm is trained to better classify the signal and background events, it may be asked to return the probability of a given event to be a signal event. We will call this an \emph{output score}. In this way, we can construct distributions of scores for signal and background events and then apply another cut on this distribution. In this case, eq.~(\ref{nevents:1}) is modified by multiplying it by another unit step function $H({\cal O}_{ML}(x_{ML},x_{ML}^c))$. The ML scores $x_{ML}$ may themselves depend on other specific parameters $\mathbf{\theta}_{ML}$ and must also be adjusted for a good performance. We discuss this in section~\ref{section:BDT}.

The most brute force method to obtain the optimal set of cuts, a multivariable scan, is also the one that is the least pragmatic. For example, the ATLAS~\cite{ATLAS14} study makes use of more than 10 kinematic variables. A hypercube in this space with just a 10-fold division in each direction represents $10^{10}$ different cut strategies. To cite another example, one can consider the search for single-top production at the Tevatron~\cite{Aaltonen:2010jr}, which trained neural networks with up to 30 variables that could be used in a cut analysis. It is evident that large grids are unfeasible without large computational facilities. 

The situation becomes even more untenable when ML algorithms are used to enhance the collider searches, since they add a much longer time of computation in the analysis chain. A deep neural network, for example, might take from several minutes to several hours to train, depending on the computational resources and the size of the training/testing samples. On the other hand, selection cuts may have a significant effect on the kinematic variables (features) which are used to train ML algorithms. These effects are often neglected but may significantly impact the performance of discrimination tools. 

Intuitively, one expects that requiring hard cuts to clean up samples would force one into a small corner of feature space where signal and background events present little distinction. This degrades the ML performance. In other words, hard cuts introduce biases which make signal and background distributions indistinguishable. Loosening the cuts reduces bias, but the gain in performance of the ML discrimination may not compensate for the increased number of background events. This, too, may lead to a degraded performance, especially when systematic uncertainties are taken into account. 

The maximum significance achievable must, therefore, be a trade-off between cuts on the kinematic variables and ML performance. We note that ML classification can be performed in two ways: (1) by generating a new distribution with the ML output classification ranking of signal and background events, where a good discriminator should give the majority of signal (backgrounds) events a score close to 1(0), for example, and subsequently using this distribution to place another cut as discussed above, and (2) using the output distributions in a multivariate statistical analysis (MVA) based on the likelihood ratio statistic for the final discrimination.

The solution to avoid expensive grid searches can be found in the data science literature itself. The most powerful ML algorithms, such as neural networks and decision trees, have a large number of parameters (called hyperparameters) which control their performance. Adjusting hyperparameters to achieve a high classification accuracy is an important goal in ML, and avoiding extensive scans in the space of hyperparameters is desirable. It is now common practice to perform either randomized grid searches or use dedicated algorithms for  model configuration~\cite{Bergstra1}. Surprisingly, a simple random search with hundreds of trials may perform as good as, or even better than, a manual search.


For large parameter spaces, however, it has been demonstrated that Bayesian optimization performs better than either manual or randomized searches~\cite{Bergstra2}. The algorithm described in ref.~\cite{Bergstra2}, implemented in the Python library \HyperOpt~\cite{url:hyperopt}, is based on the so-called sequential model-based optimization (SMBO) technique~\cite{SMBO}. This class of algorithms suggests a new model (a new configuration of parameters) at each iteration in order to optimize the criterion of Expected Improvement (EI), which is the expectation that under a model $M$ of a function $f$, $y=f(x)$ will exceed some threshold $y^c$
\begin{equation}
EI_{y^c}(x)=\int_{-\infty}^{+\infty} \max(y^c-y,0)p_M(y|x)dy
\end{equation} 
in the search for the minimum of $f$. 

The major obstacle in computing $EI(x)$ is estimating the conditional probability $p_M(y|x)$. \HyperOpt overcomes this difficulty by means of the Bayes rule, $p_M(y|x)=\frac{p(x|y)p(y)}{p(x)}$, where $p(x)$ is an assumed prior distribution of the parameters. By keeping a sorted list of observations of $y=f(x)$, it is possible to compute the quantiles $\gamma=p(y<y^c)$, while $p(x|y)$ is a non-parametric distribution estimated from previous observations along the run of the algorithm. The strategy to evaluate $p(x|y)$ in \HyperOpt is known as a Tree-structured Parzen Estimator approach, TPE for short. In TPE, $p(x|y)$ equals $\ell(x)$($g(x)$) if $y<y^c$($y\geq y^c$), thus providing an non-parametric estimate of $p(x|y)$ from previous runs of the algorithm. Further details of the algorithm can be found in ref.~\cite{Bergstra2} and references therein.

This way, it is possible to show that $EI_{y^c}(x)$ is such that 
\begin{equation}
EI_{y^c}(x)\propto \left(\gamma+\frac{g(x)}{\ell(x)}(1-\gamma)\right)^{-1}
\end{equation}
where, on each iteration, the algorithm returns the point on the parameters space $x^c$ with greatest expectation improvement.
The algorithm is efficient once $EI_{y^c}(x)$ grows as the ratio $g(x)/\ell(x)$ drops, that is, as $\ell(x)$ accumulates with the learning process and $g(x)$ represents more rare configurations. 

The main result of this section is to use  Bayesian optimization to look for better discriminating kinematic cuts. In this case, $x$ is a point in a kinematic multivariable space designed to discriminate between signal and backgrounds and $f(x)$ is an Approximated Mean Significance (AMS) function, a significance metric as defined in eqs.(\ref{sig:naive},\ref{sig:mva},\ref{sig:ams}). In section~\ref{section:joint} we will investigate an augmented searching space comprising the thresholds of the kinematic variables for cuts and the hyperparameters which models a boosted decision trees algorithm, thus performing a joint cuts plus hyperparameters search. 

\subsection{Results using Bayesian optimization in \HyperOpt}

We use \HyperOpt~\cite{url:hyperopt} for the search with the TPE strategy described above. The inputs of the program are a Python dictionary with the names and variation ranges of the variables, the prior random distributions assumed for those variables, the objective function to be minimized, and the number of experiments which the algorithm is allowed to perform in the search, that is, the number of trials. The algorithm can be easily parallelized as described in ref.~\cite{Bergstra2}, but our searches were all obtained within a single thread of the computer, thus the running time of cut searches could be greatly reduced. In appendix~\ref{app:2} we display a simple code that can be adapted by the reader for immediate use in a cut-and-count analysis.

In table~\ref{table:2} we show the kinematic variables used for cut optimization and their ranges of variation. For all of them, we assume uniform priors. The corresponding number of points in such a grid would be staggering $1.86368\times 10^{14}$ possible cut strategies.

\begin{table}[t]
\centering
\begin{tabular}{c|c}
\hline
Kinematic variable & Variation range in \HyperOpt\\
\hline\hline
$\Delta R_{ii} <$ & $(1.4,4,0.05)$ \\
\hline
$\Delta R_{ij} >$ & $(0,2,0.05)$ \\ 
\hline 
$p_T(1) >$ & $(30,100,1)$ GeV \\
\hline 
$p_T(2) >$ & $(20,70,1)$ GeV \\
\hline 
$p_{T_{ii}} >$ & $(0,200,5)$ GeV \\
\hline 
$M_{\bbaa} >$ & $(0,400,5)$ GeV \\
\hline 
$M_{b_1\gamma_1} >$ & $(0,200,5)$ GeV \\
\hline 
$|M_{\gamma\gamma}-m_h| <$ & $(5,15,1)$ GeV \\
\hline 
$|M_{bb}-m_h| <$ & $(10,30,1)$ GeV \\
\hline\hline
\end{tabular}
\caption{The kinematic variables used for cuts and their allowed variation ranges in \HyperOpt. The prior distributions for all these variables are set to uniform distributions over the ranges shown in the table within the steps shown as the last entry of each vector.}
\label{table:2}
\end{table}
Compared to the variables of table~\ref{table:1}, we also experimented with the invariant mass of the hardest $b$-jet and photon,$M_{b_1\gamma_1}$. We required the same $\Delta R$ cut for $b$-jets and photons pairs and for all $b\gamma$ combinations according to the first two rows of table~\ref{table:2}. We also put the same cut on the transverse momentum of the hardest(second hardest) photon and bottom. Of course, we could have chosen different cuts for each particle $p_T$ and $\Delta R$ pair. The rapidity cuts are kept constant throughout the experiments, $|\eta|<2.4$ for all photons and jets.

In table~\ref{table:3} we show the set of cuts that achieves the largest significance in a cut-and-count analysis found with the Bayesian search after 200 trials. The first row shows the optimized cuts and the significance, computed with $S/\sqrt{B}$, reached for the same backgrounds of ref.~\cite{Azatov:2015oxa}. In the second row we show the results for the extended backgrounds including $\ccaa$, $\bbaj$ and $\ccaj$ events. The last row displays the results for the cuts of Azatov \emph{et. al.}, ref.~\cite{Azatov:2015oxa}; the upper sub-row is the significance computed with the same backgrounds of that work, while the lower sub-row contains the $S/\sqrt{B}$ with the extended set of backgrounds considered in our work.

First of all we note that the learning process selects somewhat different sets depending on the actual size of the backgrounds. The Best (1) strategy of the first row, with smaller backgrounds, relied mainly on the $M_{\bbaa}$ and $p_{T_{\gamma\gamma}}$ variables to eliminate backgrounds. The Best (2) set of the second row, for extended backgrounds, put a stronger cut on the $\Delta R_{ii}$ compared to the Best (1) set, while the other cuts remained more or less the same. This confirms that the $\Delta R_{ii}$ variables are indeed discriminative. Second, both strategies found better discrimination putting cuts on $M_{\bbaa}$ and $p_{T_{bb,\gamma\gamma}}$ which also confirms the usefulness of these variables. Third, we observe that the optimized sets relax the $p_T$ cuts on the softer $b$'s and photons whereas strengthening the cut on the hardest particles. As in previous studies, the window around the $b\bar{b}$ peak is wider than the $\gamma\gamma$ peak. Finally, $\Delta R_{ij}$ and $M_{b_1\gamma_1}$ were found to be less important in the discrimination as observed in table~\ref{table:3}. 

\begin{figure}[!t]
\centering
\includegraphics[scale=0.35]{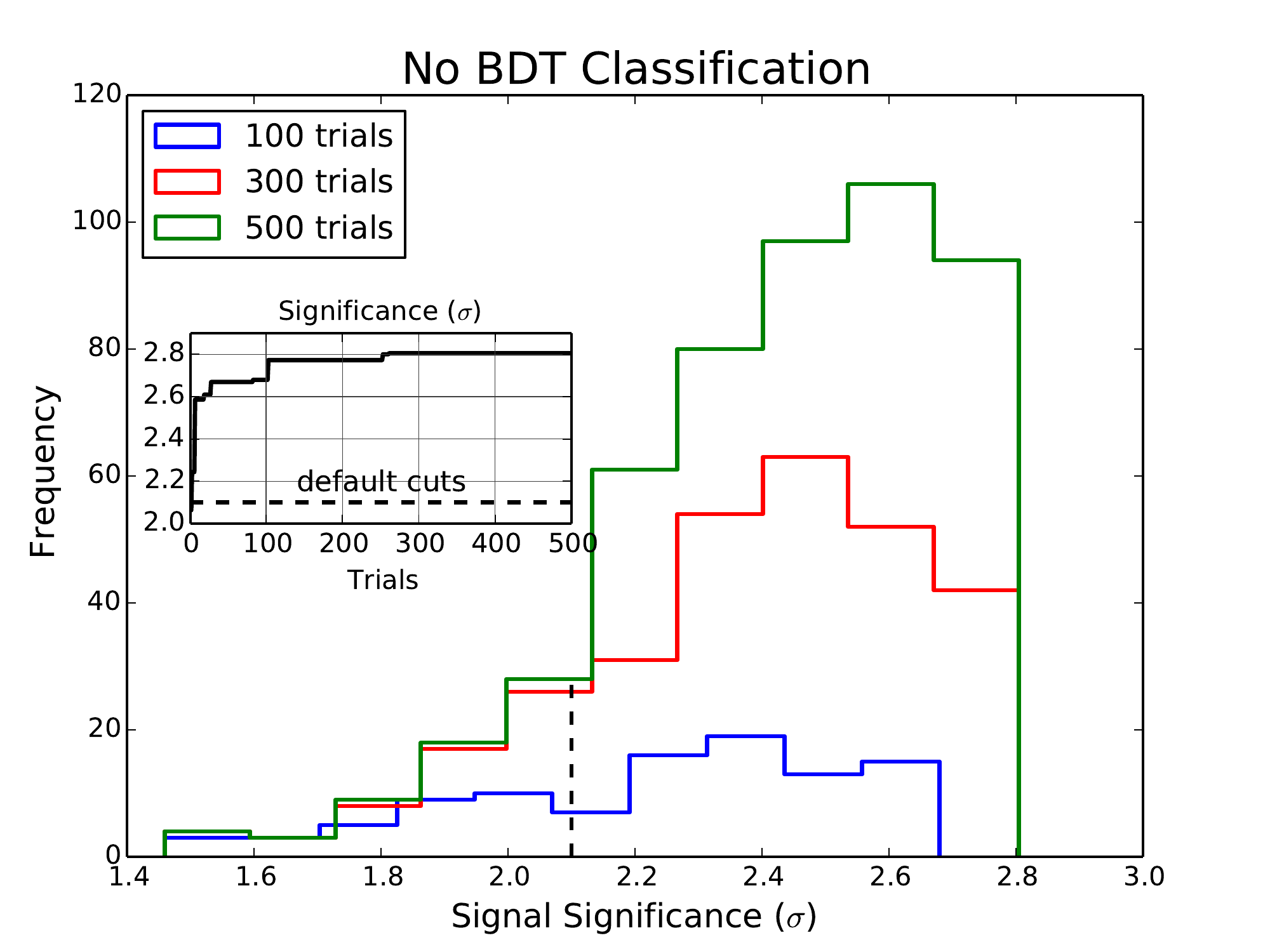}
\includegraphics[scale=0.35]{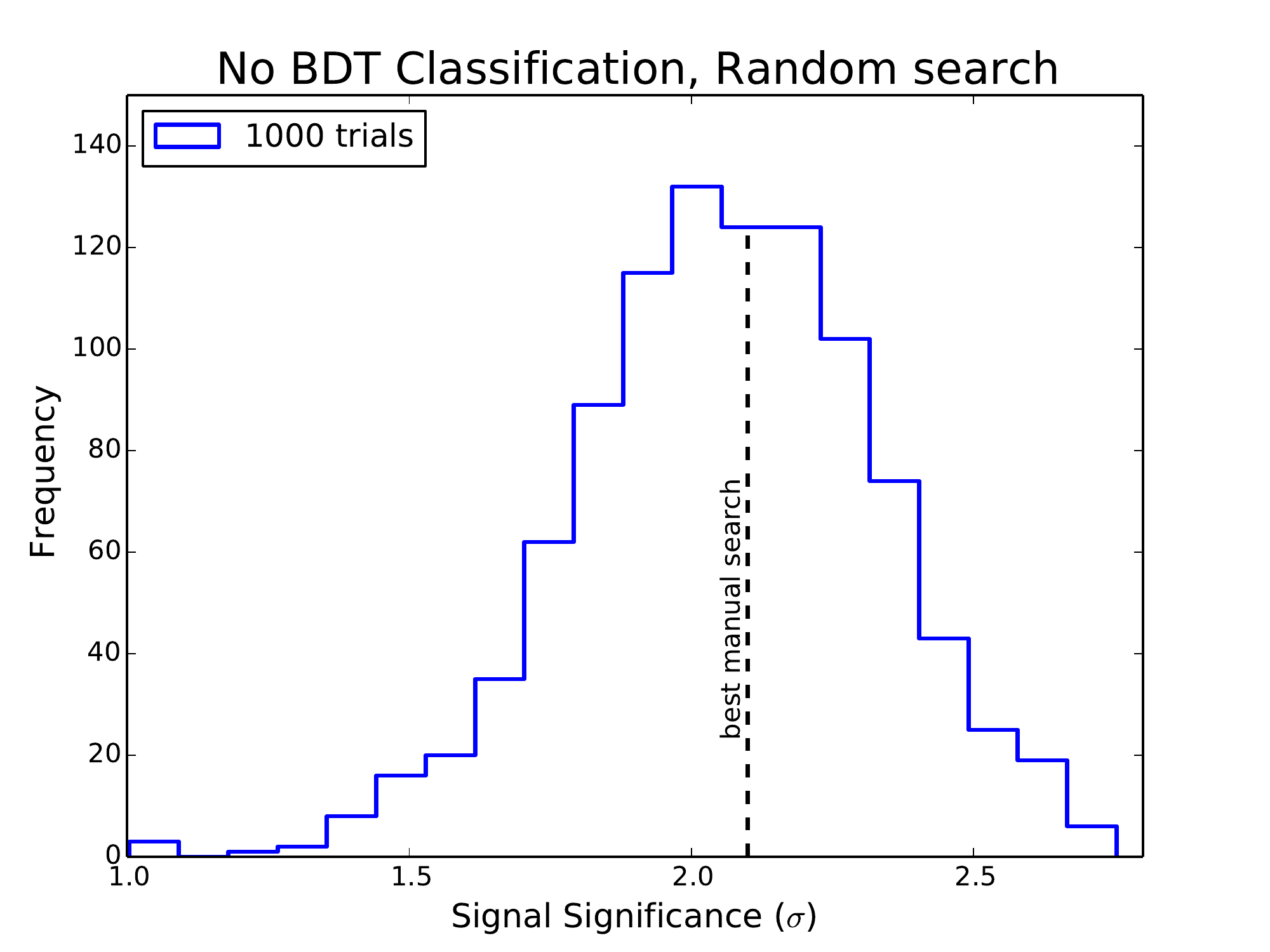}
\caption{The histograms of number of cut strategies producing a given significance interval in a cut-and-count analysis. At the left plot we show the optimized search with the TPE algorithm in \HyperOpt. The inset frame in the left plot shows the significance as a function of the number of trials. At the right plot we display a non-optimized random search after 1000 trials.
No systematics are assumed, the backgrounds are those of ref.~\cite{Azatov:2015oxa} and the $S/\sqrt{B}$ used to compute the signal significances. The black dashed line represents the results obtained with the default cuts of Azatov \emph{et. al.}, ref.~\cite{Azatov:2015oxa} in all plots.}
\label{fig:2}
\end{figure}

We now investigate how often \HyperOpt finds cuts with higher significances compared to the cuts of ref.~\cite{Azatov:2015oxa} and with the same background assumptions of that work. For this investigation we performed 500 trials and created histograms for the number of sets cuts in a given $S/\sqrt{B}$ interval as shown at the left plot of figure~\ref{fig:2}. The blue(red)[green] histogram displays the number of sets for a given AMS interval after 100(300)[500] trials. 
\begin{table}[t]
\centering
\begin{tabular}{c|c|c|c}
\hline
Kinematic cuts & $N_S$ & $N_B$ & $S/\sqrt{B}(\sigma)$ ($S/B$)\\ 
\hline \hline
 Best (1): $p_T(1) >90$ GeV, $p_T(2) >21$ GeV & & & \\
 $\Delta R_{ij}>0.65$, $\Delta R_{ii}<3.75$ & {\color{red}21.0} & {\color{red}55.1} &  {\color{red}2.81(0.38)} \\
 $M_{\bbaa}>385$ GeV, $p_{T_{ii}}>100$ GeV, $M_{b_1\gamma_1}>60$ GeV & & & \\
 $|M_{bb}-m_h|<24$ GeV, $|M_{\gamma\gamma}-m_h|<7$ GeV & & & \\ 
\hline
Best (2): $p_T(1) >86$ GeV, $p_T(2) >22$ GeV & & & \\
 $\Delta R_{ij}>0.4$, $\Delta R_{ii}<1.85$ & {\color{blue}18.0} & {\color{blue}52.1} & {\color{blue}2.48(0.35)} \\
 $M_{\bbaa}>390$ GeV, $p_{T_{ii}}>100$ GeV, $M_{b_1\gamma_1}>25$ GeV & & & \\
 $|M_{bb}-m_h|<24$ GeV, $|M_{\gamma\gamma}-m_h|<8$ GeV & & & \\ 
\hline
Default:  $p_{T}(1)>30$ GeV, $p_{T}(2)>50$ GeV & {\color{red}12.8} & {\color{red}37.1} & {\color{red}2.1(0.34)} \\
  $\Delta R_{ij}>1.5$, $\Delta R_{ii}<2$ & & & \\
  $|M_{bb}-m_h|<20$ GeV, $|M_{\gamma\gamma}-m_h|<5$ GeV & {\color{blue}12.8} & {\color{blue}48.7} & {\color{blue}1.85(0.27)} \\
\hline\hline
\end{tabular}
\caption{The rows show a set of cuts at the left column, and the number of signal and backgrounds after these cuts and the significance(signal-to-background ratio) in the subsequent columns.
The first row shows the results reached for the same backgrounds as ref.~\cite{Azatov:2015oxa} after 200 \HyperOpt trials. In the second row we show the results for the extended backgrounds including $\ccaa$, $\bbaj$ and $\ccaj$ events again with 200 Bayesian searches. The last row displays the results for the cuts of Azatov \emph{et. al.},  ref.~\cite{Azatov:2015oxa}; the upper numbers in red in this row are computed with the same backgrounds of that work, while the lower ones in blue contains are computed with the extended set of backgrounds considered in our work.}
\label{table:3}
\end{table}

Around 90\% of all Bayesian optimization searches yielded a greater significance than the 2.1$\sigma$ achieved by the cuts of Azatov \emph{et. al.}, represented by the dashed line at the left plot of figure~\ref{fig:2}. The 300 and 500 trials histograms also make evident the way the algorithm improves the objective function, $S/\sqrt{B}$ in this case. The bins of higher significances get more populated as we increase the number of trials indicating that the algorithm learns with past cut-and-count experiments in order to search for better ones as expected. This is no surprise, since the Bayesian optimization is actually a generative machine learning algorithm as described in the previous section.

In the inset frame at the left plot of figure~\ref{fig:2} we show $S/\sqrt{B}$ as a function of the number of trials. We see that after 100-200 trials, the signal significance does not change much up to 500 trials. After 200 trials, the optimized cuts achieved a significance of 2.81$\sigma$ against 2.1$\sigma$ of the manual search of ref.~\cite{Azatov:2015oxa}, a 34\% improvement. With extended backgrounds, the Bayesian search reached 2.48$\sigma$ against 1.85$\sigma$ of the cuts of ref.~\cite{Azatov:2015oxa}, again roughly the same improvement. A larger $S/B$ was also achieved as shown in table~\ref{table:3}. 
%

We also point out that the previous works of refs.~\cite{Kling:2016lay,Cao:2016zob,Cao:2015oaa} approached in different ways the optimum locus of the variables space for better discernment between signal and backgrounds. Similarly to our findings, those works also highlight the relative importance of the $M_{\bbaa}$, $p_{T_{\gamma\gamma}}$, $p_{T_{bb}}$, and $\Delta_{\gamma\gamma,bb}$ variables. 

\subsection{Reliability of the Bayesian Search}

In order to probe the reliability of the Bayesian approach, we performed an exhaustive grid search in a reduced variables space. We choose the 4-dimensional $(\Delta R_{ii},\; \Delta R_{ij},\; M_{\bbaa},\; p_T(1))$ space with ten evenly spaced values in each direction amounting to $10^4$ different sets of cuts.  We allowed \HyperOpt to carry out up to 300 TPE trials. Both $\Delta R$ ranges were chosen to lie in $(1,3,0.2)$, the $\bbaa$ invariant mass, $(300,600,30)$ GeV, and the $p_T(1)$ variable, $(20,70,5)$ GeV.

The maximum $S/\sqrt{B}$ found were
\begin{eqnarray}
& \hbox{{\bf Grid search:}} & 2.11\sigma,\; \Delta R_{ii}<1.6,\; \Delta R_{ij}>1.0,\; M_{\bbaa}>390\;\hbox{GeV},\; p_T(1)>25\;\hbox{GeV} \nonumber \\
& \hbox{{\bf Optimized search:}} & 2.06\sigma,\; \Delta R_{ii}<1.6,\; \Delta R_{ij}>1.8,\; M_{\bbaa}>390\;\hbox{GeV},\; p_T(1)>25\;\hbox{GeV}\nonumber \\
\end{eqnarray}

The only different cut was in the less discriminative variable $\Delta R_{ij}$, for the all the other ones, Bayesian optimization was able to find the same cut thresholds of the Grid search. Of course, in a much larger searching space it is hard to tell how close to the best grid point the Bayesian optimization gets, but our results show that the cut strategies found with hundreds of trials improve significantly the statistical significances compared to the manual searches of table~\ref{table:1}. We also point out that other open source algorithm optimization programs are available~\cite{spearmint} for experimentation.

\subsection{Random {\emph versus} Manual Search}

In phenomenological analyses, one frequently tunes the cut thresholds by visually estimating the regions of variable space which are more populated by signal or background events. Sometimes, after a first round of requirements, one looks for more discriminative variables to apply cuts on. The entire process, however, is not optimized. The similar results found by manual searches of this nature, for the cut thresholds displayed in table~\ref{table:1}, suggest that the majority of cut strategies should indeed perform nearly identically by this method.

Another strategy to avoid large grid scans is simply performing a random search for cuts. As we discussed in the previous section, this approach presents good results in the search for ML hyperparameters according to ref.~\cite{Bergstra1}. In order to investigate how the manual strategies compare to a random search, we allowed for 1000 trials in \HyperOpt, running in the random mode (see appendix~\ref{app:2} for more details), in the variables region of table~\ref{table:2}. We then computed $S/\sqrt{B}$ for each set of cuts without systematics and with the same backgrounds of ref.~\cite{Azatov:2015oxa}. The search lasted around 20 minutes with a single thread.

At the right plot of figure~\ref{fig:2} we show the histogram of the number of cut strategies for a given significance interval in this random search. The vertical dashed line is the significance of 2.1$\sigma$ reached by the best manual search of ref.~\cite{Azatov:2015oxa}. The mean of the distribution is 2.06$\sigma$ with $0.27$ standard deviation. Around 45\% of all cut strategies result in a signal significance larger than 2.1$\sigma$. In other words, a good manual search is likely to reproduce just the mean performance of a random search when we look for a promising region of the variables space for cut-and-count. We suspect that similar behavior can be observed in other phenomenological analysis based on cut-and-count. 

As observed in ref.~\cite{Bergstra1}, the Bayesian search performed slightly better than the random search in our case too. However, while a thousand experiments were necessary to reach an $\sim 2.7\sigma$ of significance in the random search, with just 200 trials is possible to reach around $2.8\sigma$ as we see in figure~\ref{fig:2}. Both searches, however, present an enhancement compared to the manual searches of table~\ref{table:1}.

We now investigate how the Bayesian cut optimization works when systematic uncertainties are present.

\subsection{Optimization with Systematic Uncertainties}
\label{section:systematics}
 As we observed in the previous section, the optimization procedure is able of not just increasing the signal significance but also the $S/B$ ratio which is essential when we take systematic uncertainties into account in the statistical analysis. This observation leads us to investigate whether \HyperOpt would also be able to find cuts with higher $S/B$ in order to tame the systematics.

\begin{figure}[!t]
\centering
\includegraphics[scale=0.5]{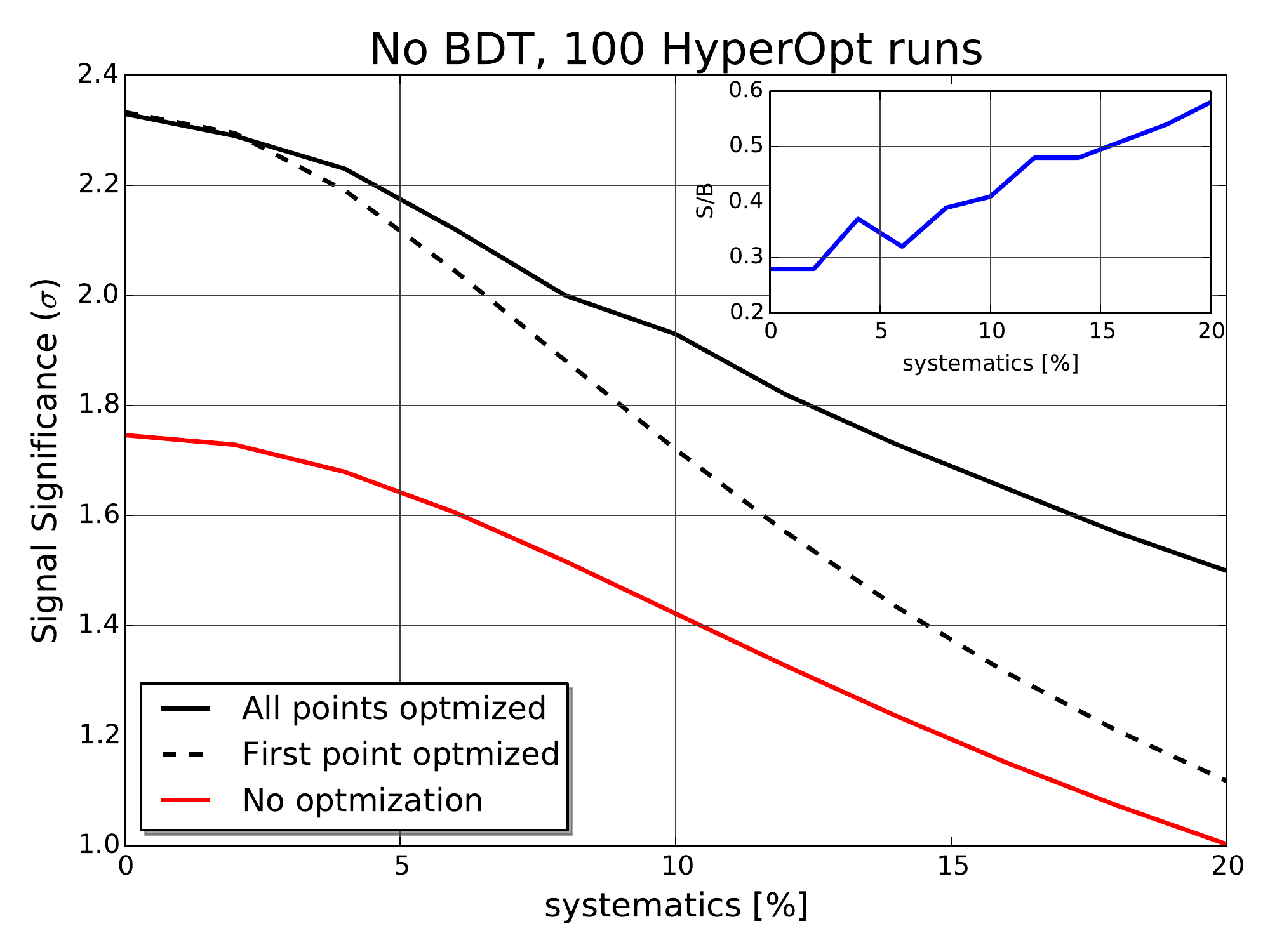}
\caption{The $S/\sqrt{B+(\varepsilon_B B)^2}$ significance metric as function of $\varepsilon_B$, the systematic uncertainty in the total background rate. The red line represents the default cuts of Azatov \emph{et. al.}, ref.~\cite{Azatov:2015oxa}, the black dashed assumes an optimized strategy just for the 0\% systematics point, while for the solid upper line, the algorithm was solicited to learn the best cuts for each systematics level from 0 to 20\%. In the inner plot we show the $S/B$ ratio for the point-to-point optimization case.}
\label{fig:3}
\end{figure}

In figure~\ref{fig:3} we show the signal significance in terms of the background rate systematic uncertainty $\varepsilon_B$ from 0 to 20\% after 100 trials. The red solid line represents the significance for the default cuts of Azatov \emph{et. al.} The points of the black dashed line are obtained by optimizing only the cuts of the 0\% case and then using $S/\sqrt{B+(\varepsilon_B B)^2}$ to extrapolate the significance for other $\varepsilon_B$, keeping the same set of cuts found in the no systematics case. This is not the best that can be done, though, as the $S/B$ ratio remains the same as in the no systematics scenario. The upper black solid line shows the results when we optmize the significance function for each systematics level. In this case, the Bayesian algorithm is able to find points with larger $S/B$ ratio trying to overcome the systematics constraints. The inset plot shows that \HyperOpt learned that $S/B$ should double from the 0 to the 20\% systematics case to reach larger significances.

\begin{figure}[t]
\centering
\subfigure[\label{fig01}]{\includegraphics[scale=0.3]{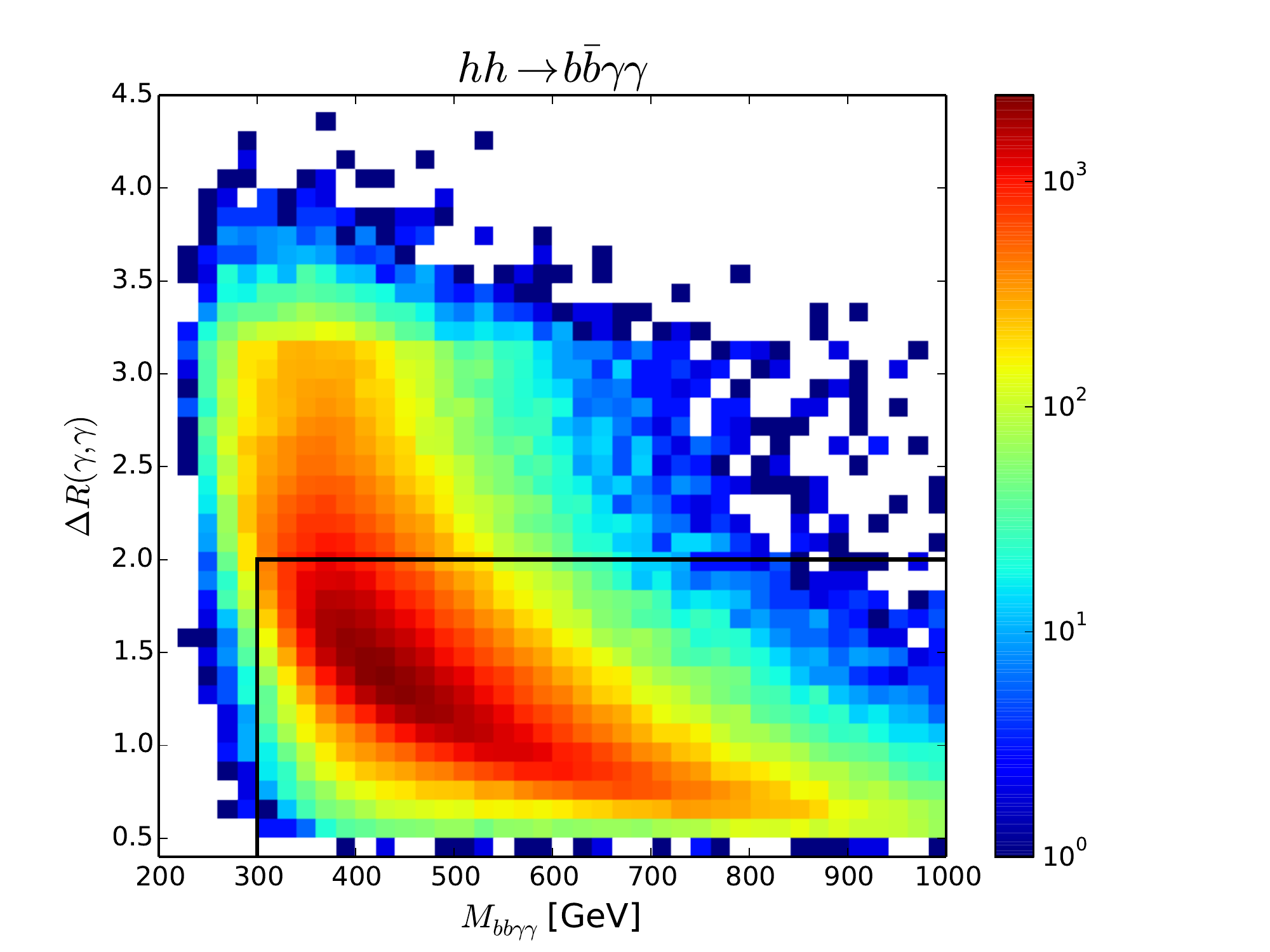}}
\subfigure[\label{fig02}]{\includegraphics[scale=0.3]{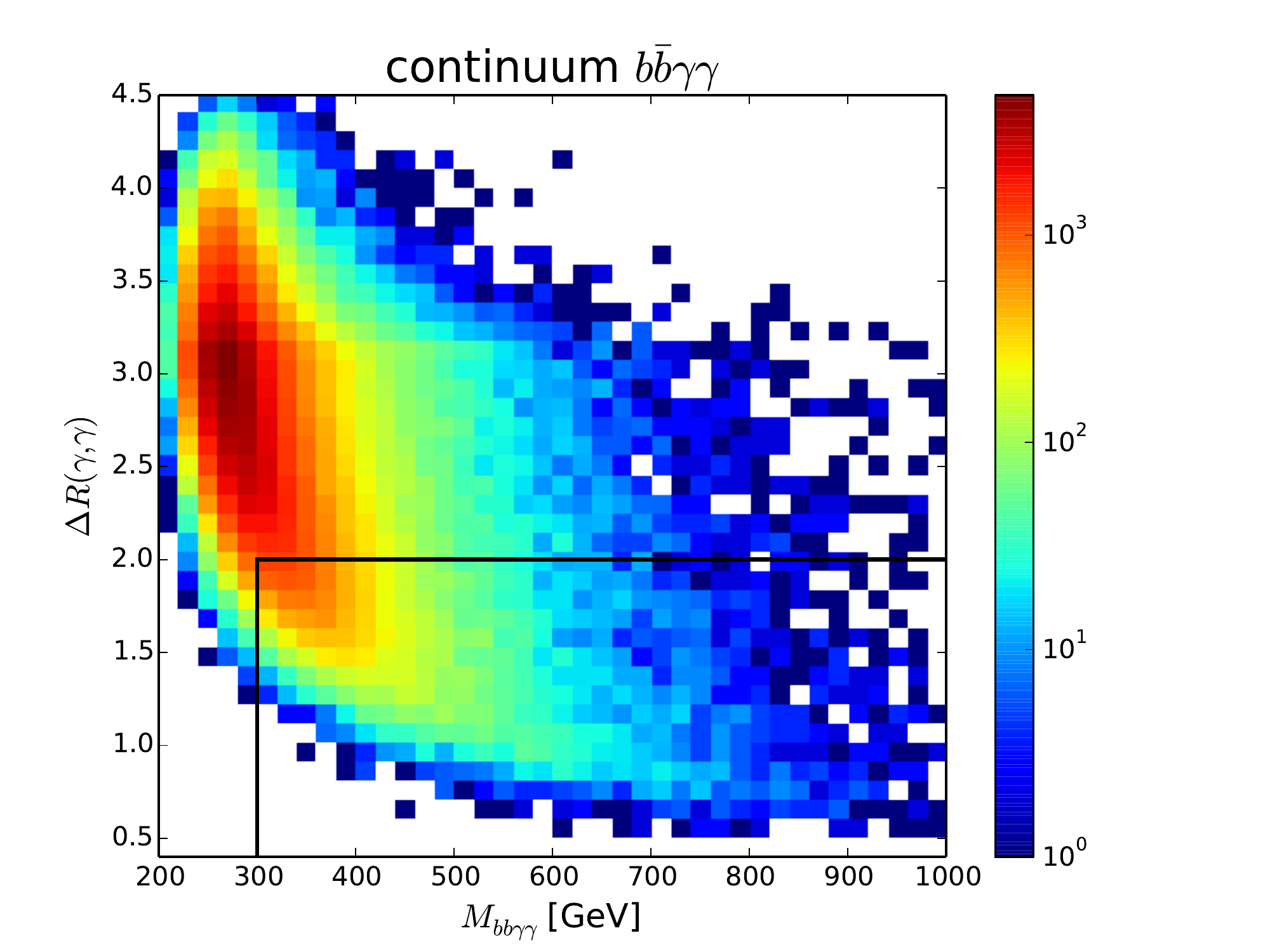}}\\
\subfigure[\label{fig01}]{\includegraphics[scale=0.3]{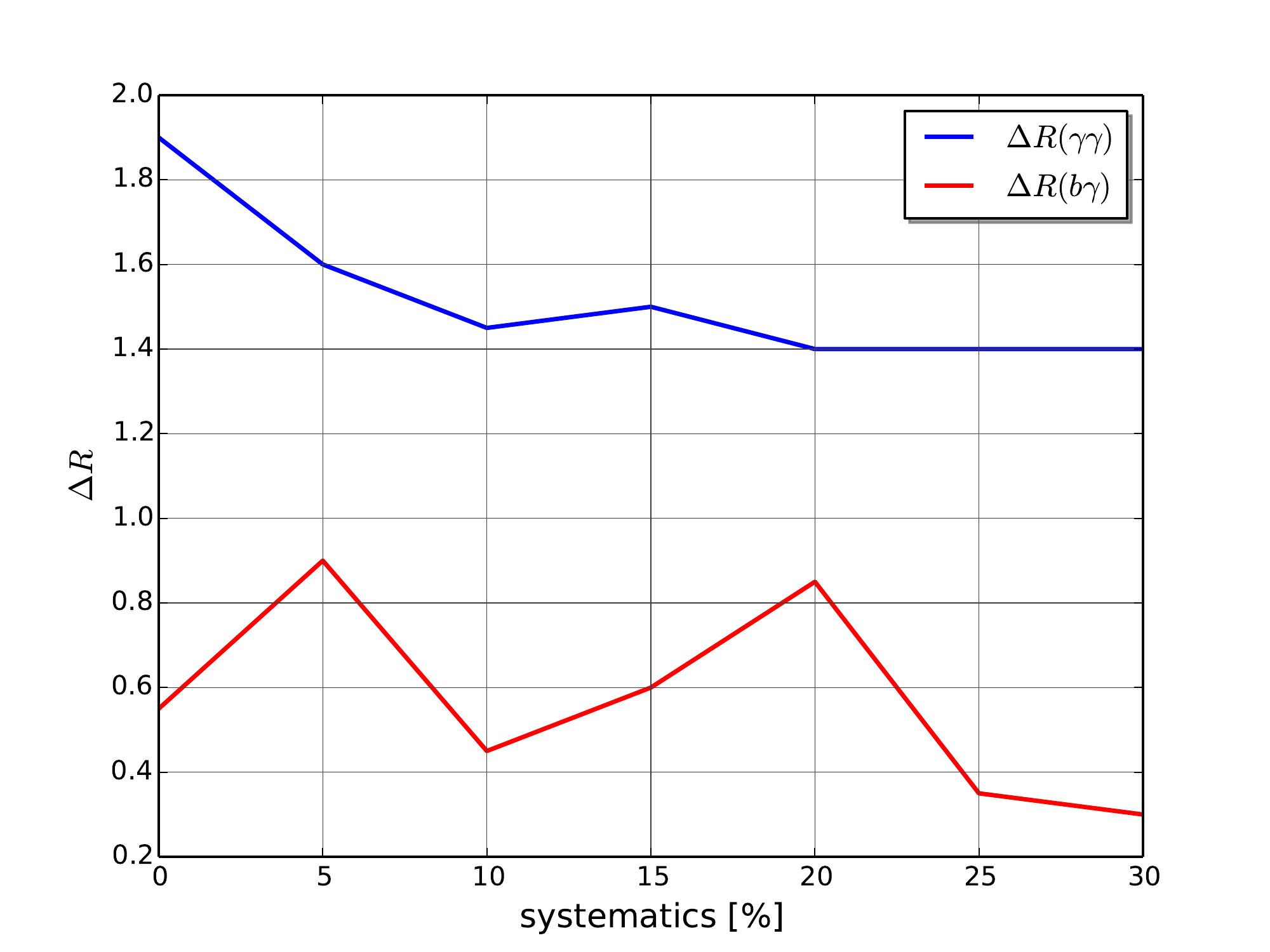}}
\subfigure[\label{fig02}]{\includegraphics[scale=0.3]{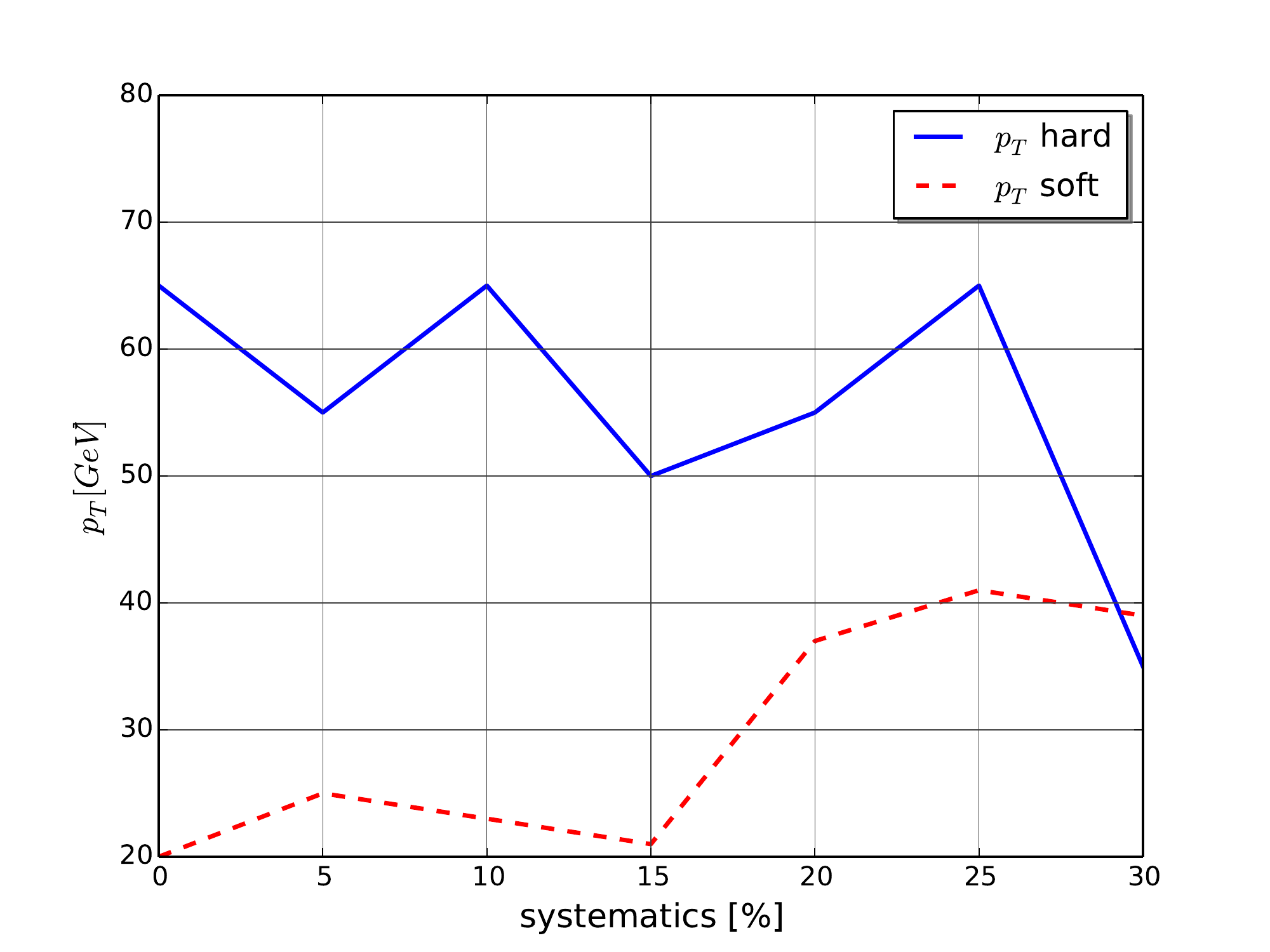}}\\
\subfigure[\label{fig03}]{\includegraphics[scale=0.3]{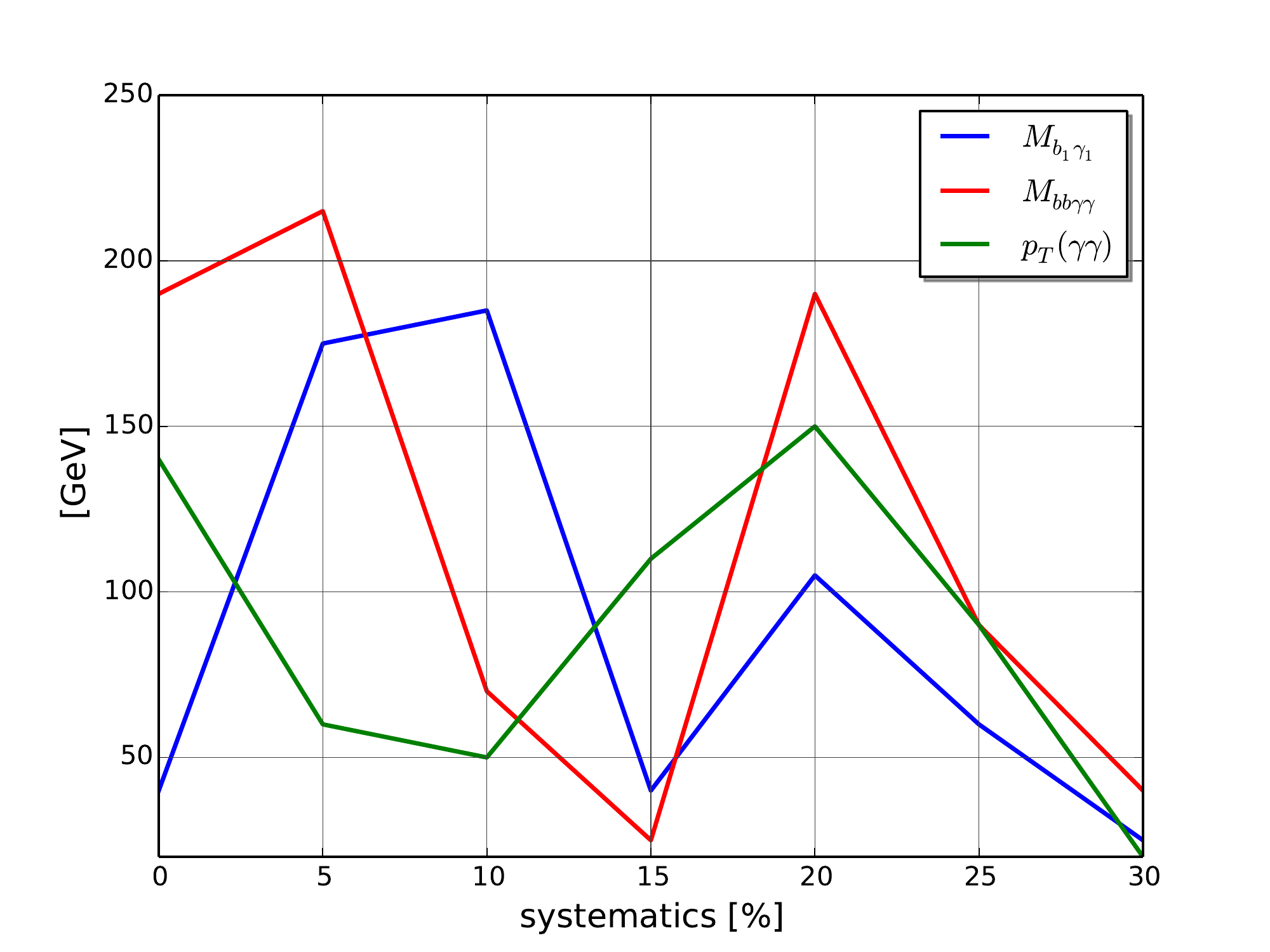}}
\subfigure[\label{fig04}]{\includegraphics[scale=0.3]{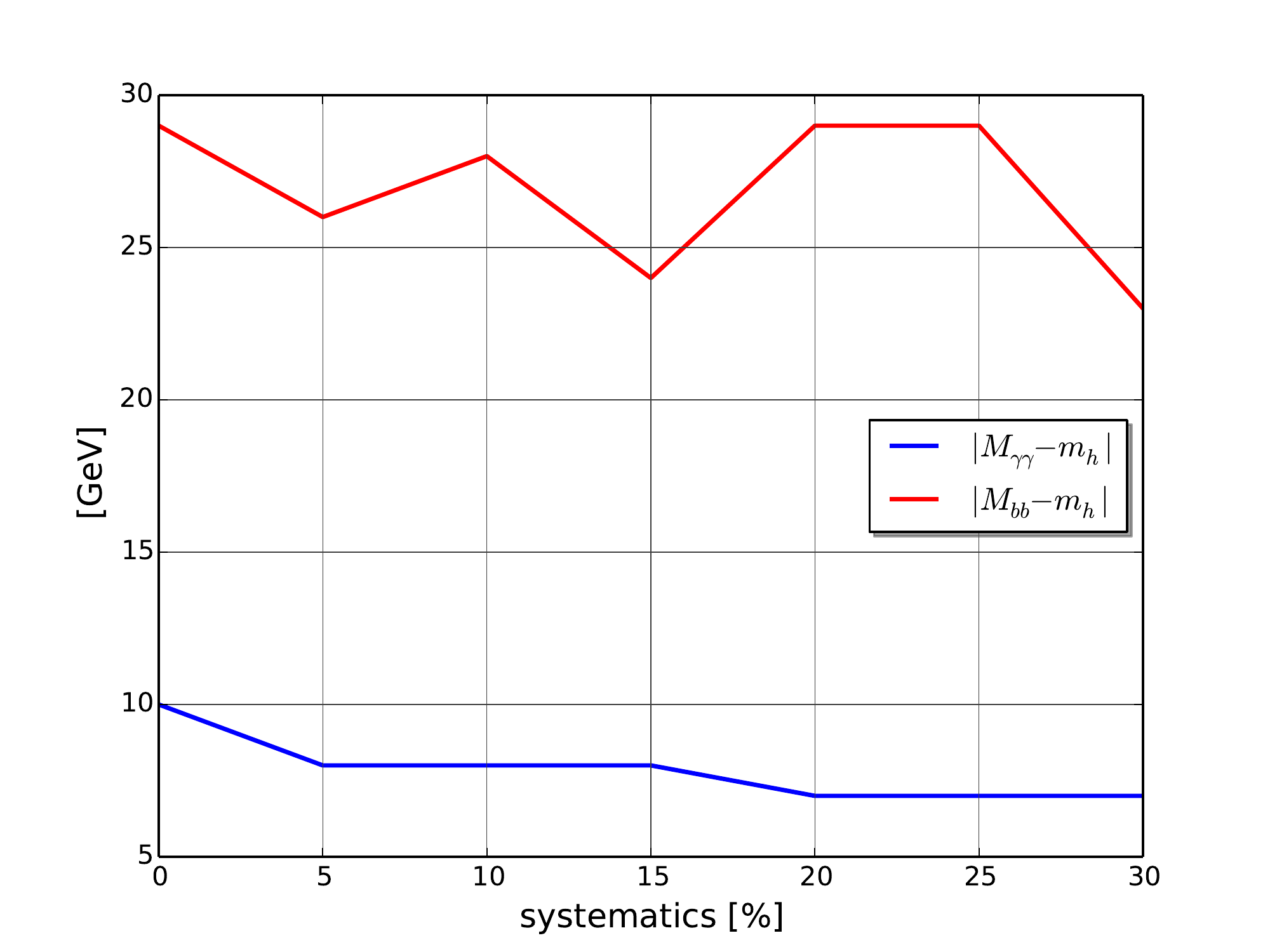}}
\caption{The learning process evolution of the cut variables in the search for maximum significance in the presence of increasing systematic uncertainties are displayed in panels (c-f). Panels (a) and (b), show the correlation between two of the most discriminative kinematic variables, $\Delta R_{\gamma\gamma}$ and $M_{\bbaa}$ for the the signal and the dominant $\bbaa$ background, respectively. In the panel (c), we show the distance of pairs of particles in the $(\eta,\phi)$ plane. The panel (d) displays the transverse momentum of the hardest and second hardest $b$'s and photons. In the panels (e) and (f), various invariant mass combinations used in the discrimination plus the transverse momentum of the pair of photons.}
\label{fig:4}
\end{figure}

As a consequence of larger $S/B$ ratios, the difference between the point-to-point optimized significance and the 0\% optimized curves gets larger as the systematics increase. It is also interesting to observe how the algorithm learns to increase the signal-to-background ratio and reach high significances as the systematics get more important. For that aim we show in figure~\ref{fig:4} the cut thresholds of some key variables used in the analysis with systematics up to 30\%. 

Some clear tendencies are noticeable: the preferred variables to hardening cuts are $\Delta R_{ii}$, the window around the $bb$ and $\gamma\gamma$ mass peaks, especially this last one, and the transverse momentum of the softer particles shown in panels (c), (f) and (d) of figure~\ref{fig:4}, respectively. On the other hand, $\Delta R_{ij}$ and the transverse momentum of the harder particles become less relevant. We already knew that $\Delta R_{ij}$ is not so important for the discrimination as the other variables. The softening of the cut of $p_T(hard)$, however, can be understood in view that we are not trying to optimize $S/B$ but the significance metric, and the algorithm seems to find a way through the second hardest $p_T$ cut instead. Despite being more erratic, a tendency to irrelevance is also observed in other discriminants like $M_{\bbaa}$, $M_{b_1\gamma_1}$ and $p_{T_{\gamma\gamma}}$, for example, as seen in panel (e) of figure~\ref{fig:4}. This can be explained in view of the panels (a) and (b) which show the correlation between two of the most discriminative variables, $\Delta R_{\gamma\gamma}$ and $M_{\bbaa}$. A hard cut on $\Delta R_{\gamma\gamma}$ makes a cut on $M_{\bbaa}$ somewhat irrelevant and vice-versa. Of course, this does not mean that this is the only way to increase $S/B$, but it does suggest that not all the kinematic variables are relevant for that task at the same time.

We especially note that for the level of background rate systematics estimated by the ATLAS and CMS Collaborations, around 10\%~\cite{ATLAS14, ATLAS17, CMS}, the optimized cuts give a  significance of 1.9$\sigma$ against $\sim 1.4\sigma$ of the default cuts of Azatov \emph{et. al.}, all with the extended backgrounds.

\section{Signal \emph{versus} Background Discrimination with Boosted Decision Trees}
\label{section:BDT}

The analysis presented in the previous section has been based solely on cut-and-count and can be employed in any phenomenological study where optimal cuts are necessary to clean up backgrounds and raise the signal significance. In appendix~\ref{app:2} we give more details about implementing this procedure in a simple and fast Python code.

In this section, we go beyond the cut-and-count analysis and focus exclusively on proposing tools to obtain even larger significances in the search for double  Higgs production at the LHC,  with and without systematics. Our goal now is to show that training a Boosted Decision Tree (BDT) algorithm to better classify signal and backgrounds events, in addition to the procedure of using optimal cuts to select the best volume of the features space for the BDT training, increases the signal significance dramatically.

We present our results in three stages. In section~\ref{section:BDTcuts}, we first introduce the kinematic observables used in the BDT analysis, and provide a discussion of the interplay between BDT classifiers and cut selections, without addressing the question of cut optimization. In section~\ref{section:sequential} we sequentially optimize the cuts on the kinematic observables using \HyperOpt, and then optimize the BDT hyperparameters. Finally, in section~\ref{section:joint}, we perform a joint optimization of the kinematic cuts and the BDT hyperparameters. 

\subsection{BDT Analysis Without Cut Optimization}
\label{section:BDTcuts}

The performance of any ML algorithm aimed to better classify signal and background events, or even an MVA analysis based on likelihood ratios, depends strongly on the portion of the feature space from which the events are selected, in other words, the number of signal and background are as follows
\begin{equation}
N_{ev}=N_{ev}(\{x_k^c,k=1,\cdots, n_c\}\cup\{x^c_{ML}(\theta_{ML},\{x_k^c,k=1,\cdots, n_c\})\}\; ,
\end{equation}
where $\theta_{ML}$ represents the hyperparameters of the ML algorithm and $N_{ev}=S(B)$ is the number of signal(total background) events. This is especially true in subtle searches for new physics, and is the reason we have investigated the Bayesian optimization method thoroughly in the previous section.

Ideally, the least biasing portion of any variables space is the one with minimal cuts, possibly requiring just acceptance and trigger cuts. However, in processes with low signals and large backgrounds like $pp\to\bbaa$, if one  employs just acceptance cuts, detection efficiencies and even takes $b$-tagging into account, one is still presented with backgrounds that are many orders of magnitude larger than the signal. This would require a ML classifier with an extremely exquisite signal acceptance \emph{versus} background rejection performance, which cannot be reached in practice. On the other hand, applying harder cuts may not necessarily degrade the ML performance to the point of making them useless for further discrimination. 

Therefore, a trade-off between cuts and ML performance should be expected in a phenomenological analysis. We now proceed to study this interplay.

\begin{figure}[!t]
\centering
\subfigure[\label{fig01}]{\includegraphics[scale=0.35]{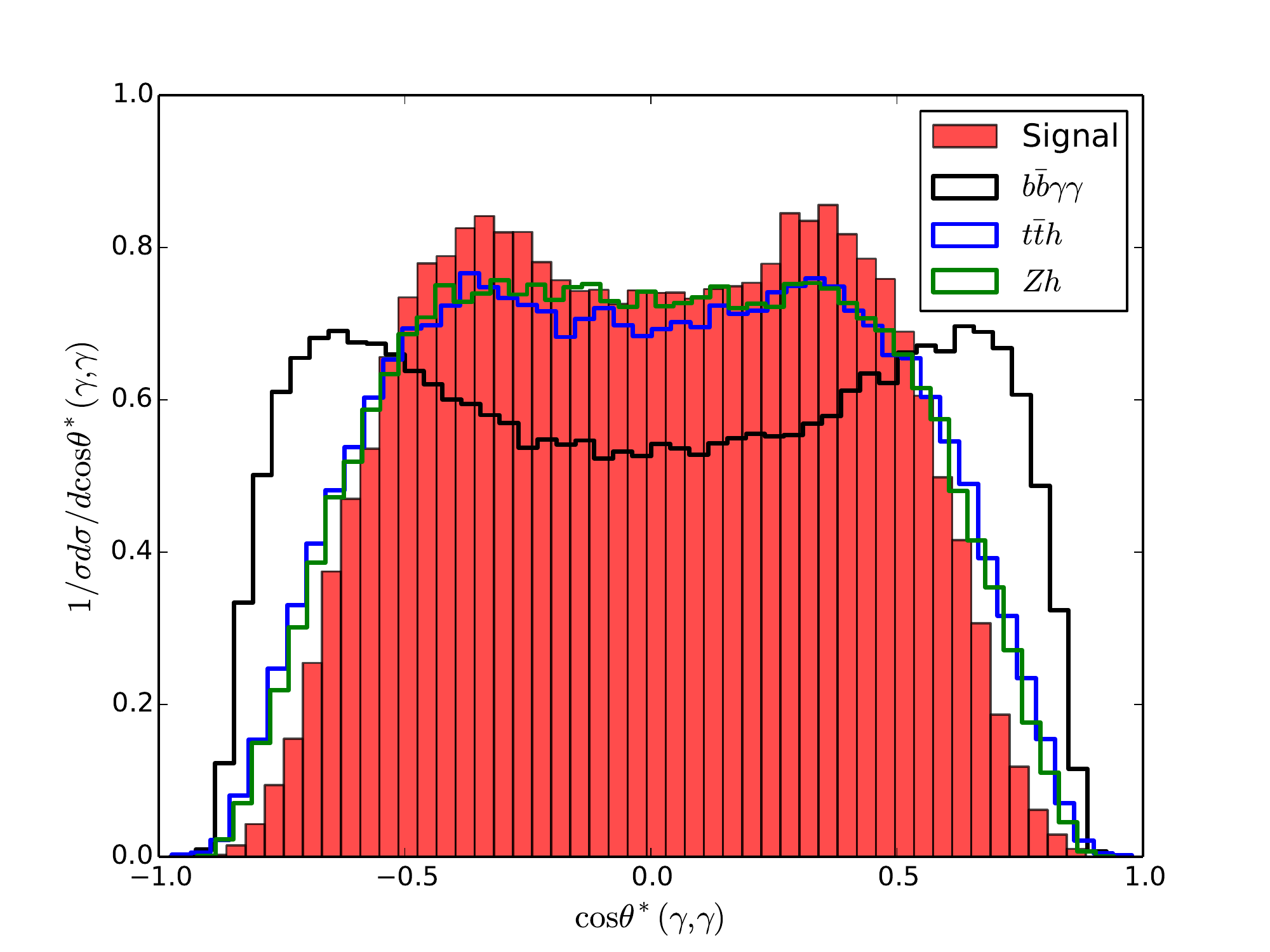}}
\subfigure[\label{fig02}]{\includegraphics[scale=0.35]{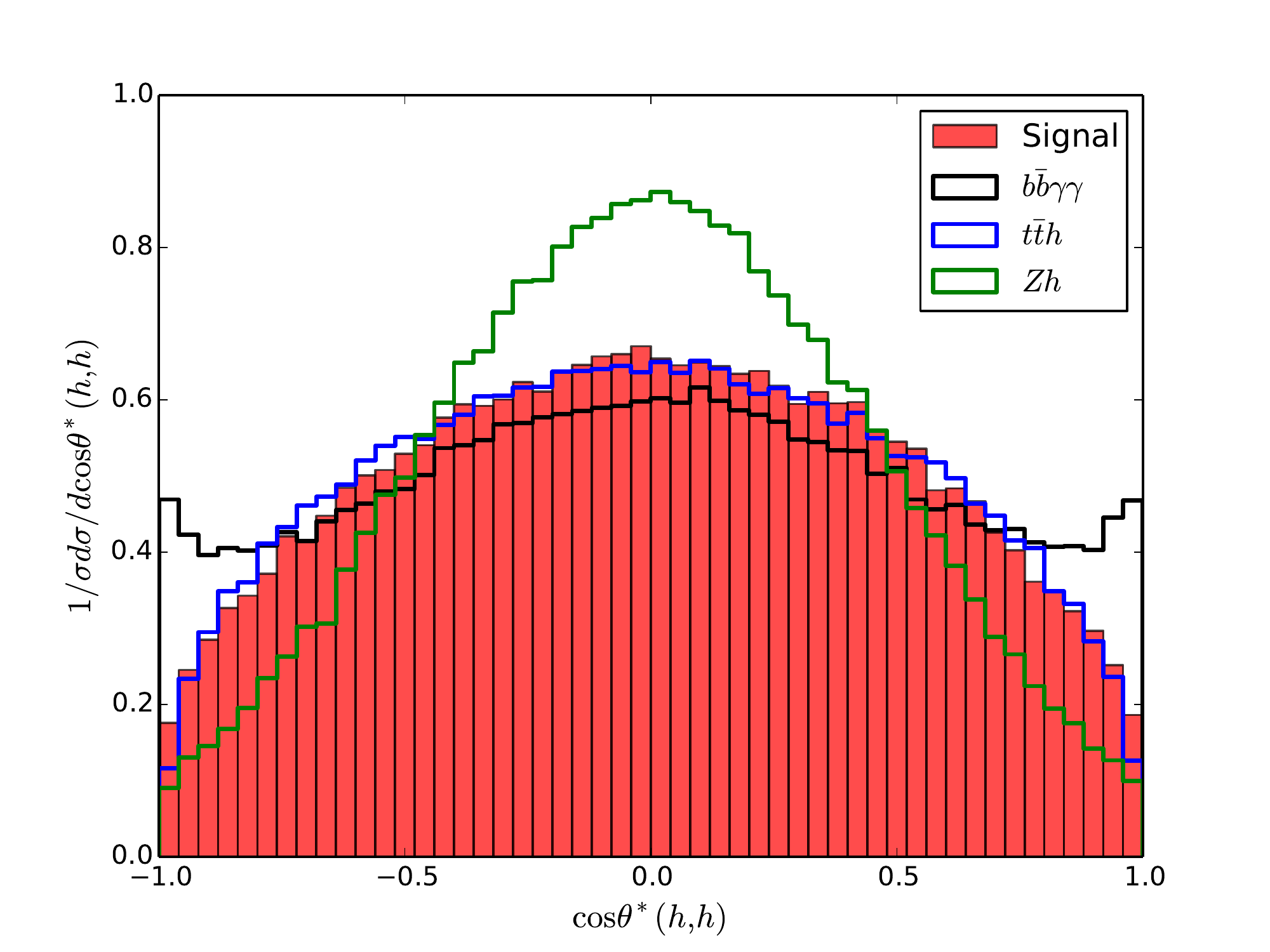}}\\
\subfigure[\label{fig03}]{\includegraphics[scale=0.35]{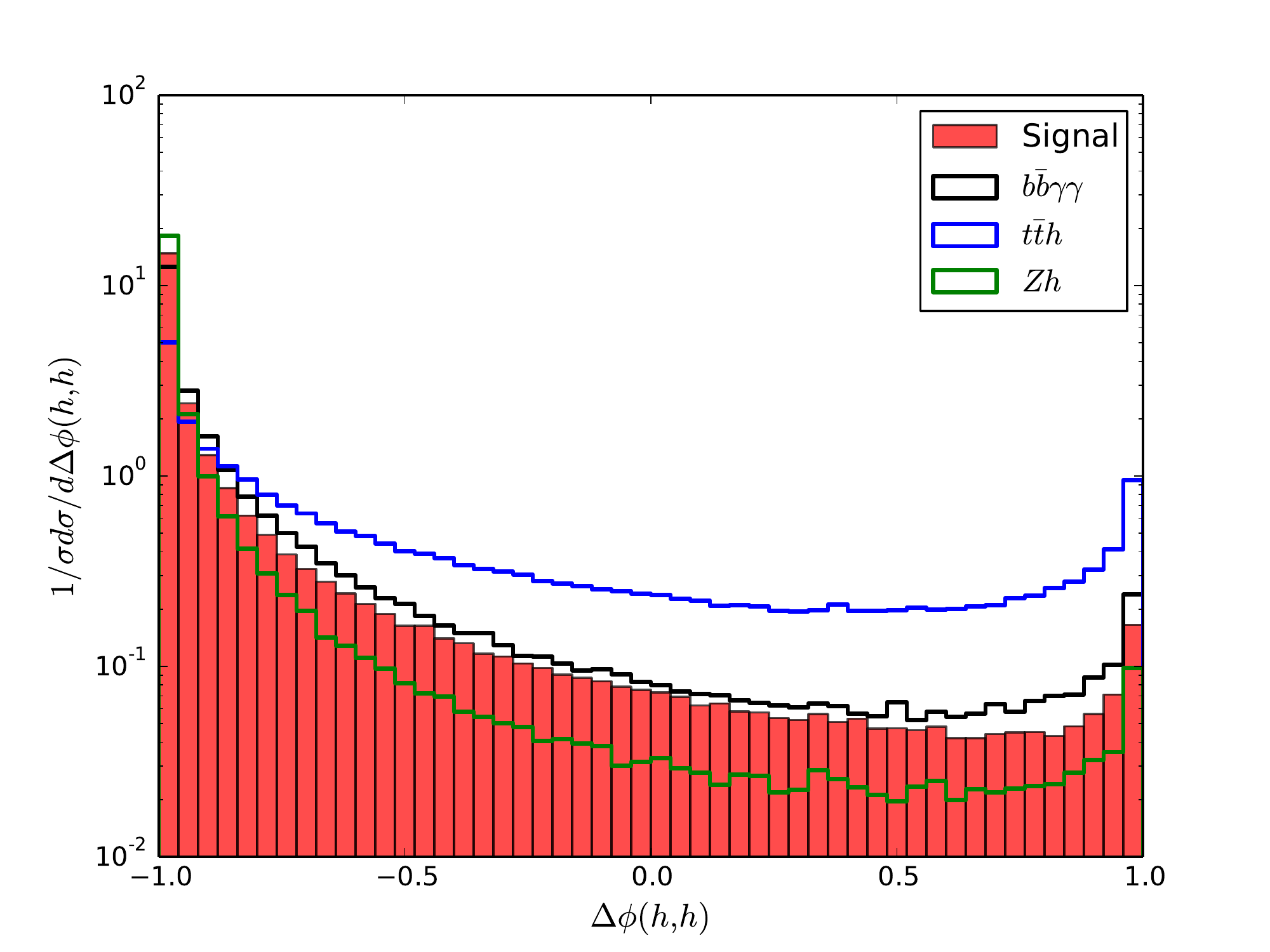}}
\subfigure[\label{fig04}]{\includegraphics[scale=0.35]{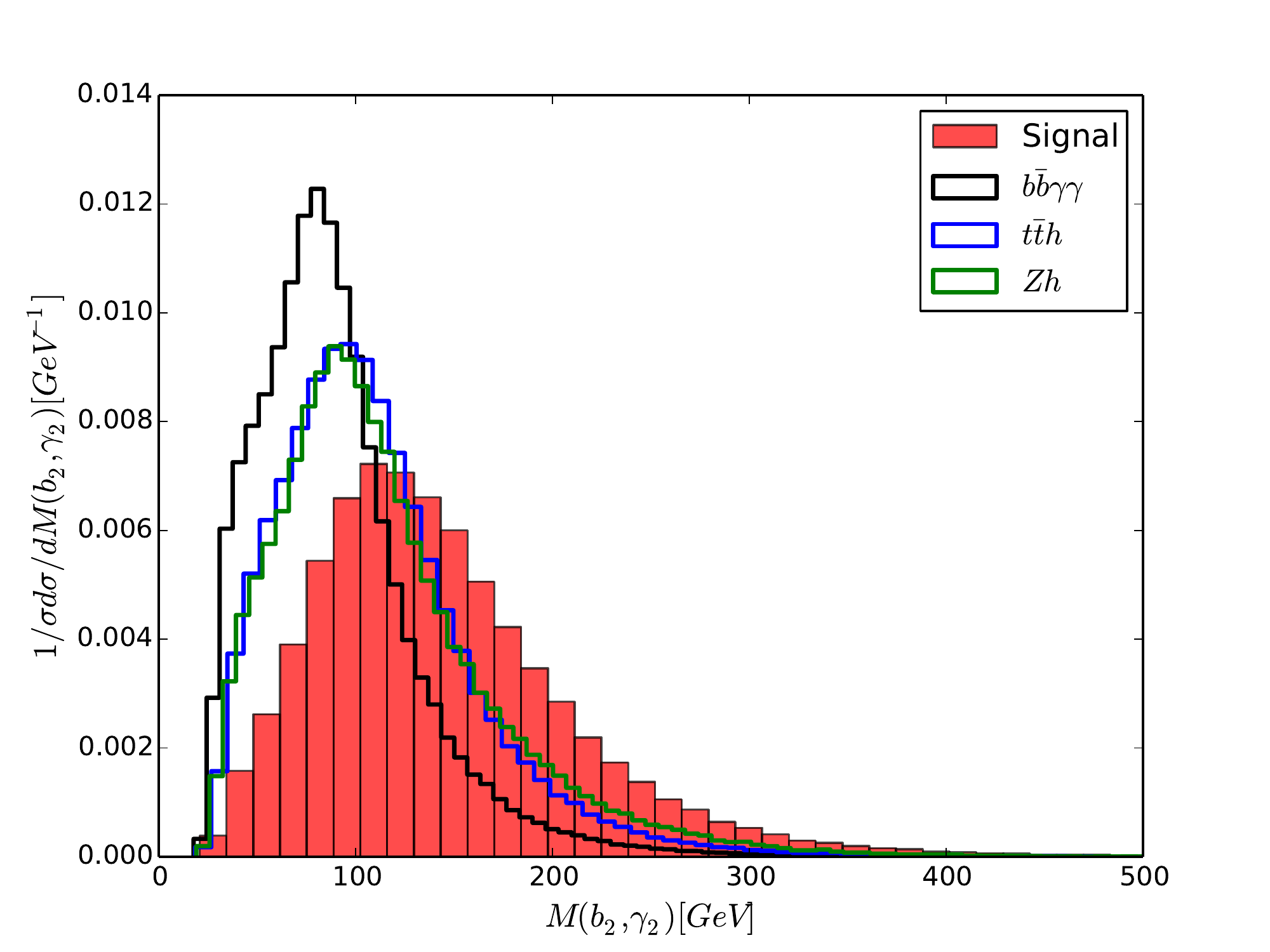}}
\caption{Four out of the 27 kinematic distributions of signal (shaded red), and the backgrounds $b\bar{b}\gamma\gamma$ (black), $t\bar{t}h$ (blue) and $Zh$ (green), used for the BDT discrimination. In (a), we show the Barr variable of the two photons (see the text for its description). Plots (b) and (c) display the Barr variable and the difference of the azimuthal angle of the reconstructed Higgs pair, respectively. In panel (d), the invariant mass of the second hardest $b$-jet and photon.}
\label{fig:5}
\end{figure}
%

We use the \xgb~\cite{TChen} implementation of BDTs for Python for its very good discrimination performance, speed and capacity of parallelization. The events features used to train the BDT are as follows:
\begin{enumerate}
\item transverse momentum of the two hardest $b$-jets and photons: $p_T(b_1,b_2)$ and $p_T(\gamma_1,\gamma_2)$
\item transverse momentum of $b\bar{b}$ and $\gamma\gamma$ pairs: $p_T(bb)$ and $p_T(\gamma\gamma)$
\item invariant mass of all four combinations of a $b$-jet and a photon: $M_{b_i\gamma_j},\; (i,j)=1,2$ 
\item invariant mass of the two $b$-jets and two photons of the event: $M_{\bbaa}$
\item distance between pairs of bottoms and photons: $\Delta R(bb)$, $\Delta R(\gamma\gamma)$ and all the four combinations of a $b$-jet and a photon $\Delta R(b_i\gamma_j),\; (i,j)=1,2$
\item the Barr variable~\cite{Barr:2005dz,Alves:2007xt} between all the six combinations of two particles in the event defined as
$\cos\theta^*_{ij}=\tanh\left(\frac{\Delta\eta_{ij}}{2}\right)$ where $\Delta\eta_{ij}$ is the rapidity separation of the $i$ and $j$ particles
\item the Barr variable between the two reconstructed Higgs bosons, $\cos\theta^*_{hh}$
\item azimuthal angle difference between the two reconstructed Higgs bosons, $\Delta\phi(h,h)$
\item missing energy of the event
\item the number of charged leptons with $p_T > 20$ GeV and $|\eta|<2.5$
\end{enumerate}

These are 27 features in total. We do not use all of them for kinematic cuts; just those shown in the first and second rows of table~\ref{table:2}. The missing energy and the number of charged leptons are used to better distinguish the multi-jet backgrounds and semi-leptonic $\tth$ backgrounds. In figure~\ref{fig:5} we show some other good features besides the ones shown in figure~\ref{fig:1}. We simulated $\sim 240000$ signal and $\sim 640000$ background events to train, test and cross-validate the BDTs. After optimized cuts we observed that the number of Monte Carlo samples of signal and background events get much more balanced.

We preprocess the features prior to the BDT training which improves their performances. First, to the distributions with skewness larger than 1.0 we add a small value of $10^{-8}$, the logarithm is taken and then they are normalized as in ref.~\cite{Baldi:2014pta}. All the features are rescaled to smaller and standardized ranges better suited for the training process.

The behavior of the statistical significance in terms of the output scores may sometimes oscillate very badly if the number
of test samples is small as a consequence of not too smooth signal and background scores distributions~\cite{HiggsML}. We checked that the AMS function, in terms of the score cut threshold for one of the five evaluations of the BDT in the five-fold cross validation for an optimized set of cuts, is very smooth and well behaved. The maximum AMS, in this case, occurs for scores cut around 0.5. For all the cut strategies with BDTs, the threshold score is chosen in order to achieve the maximum significance. 

\begin{figure}[!t]
\centering
\subfigure[\label{fig01}]{\includegraphics[scale=0.35]{dRa1b1.pdf}}
\subfigure[\label{fig02}]{\includegraphics[scale=0.35]{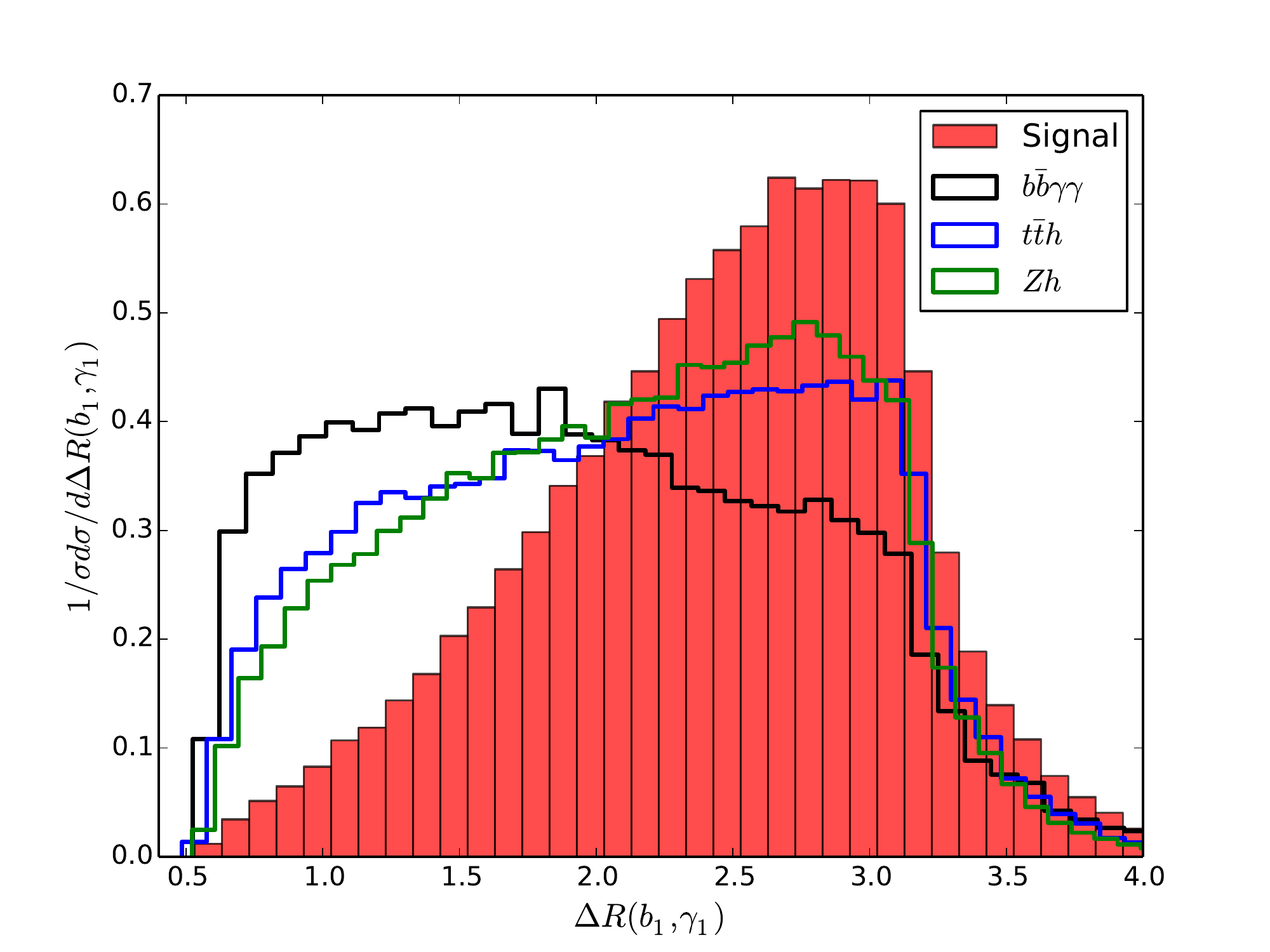}}\\
\subfigure[\label{fig03}]{\includegraphics[scale=0.35]{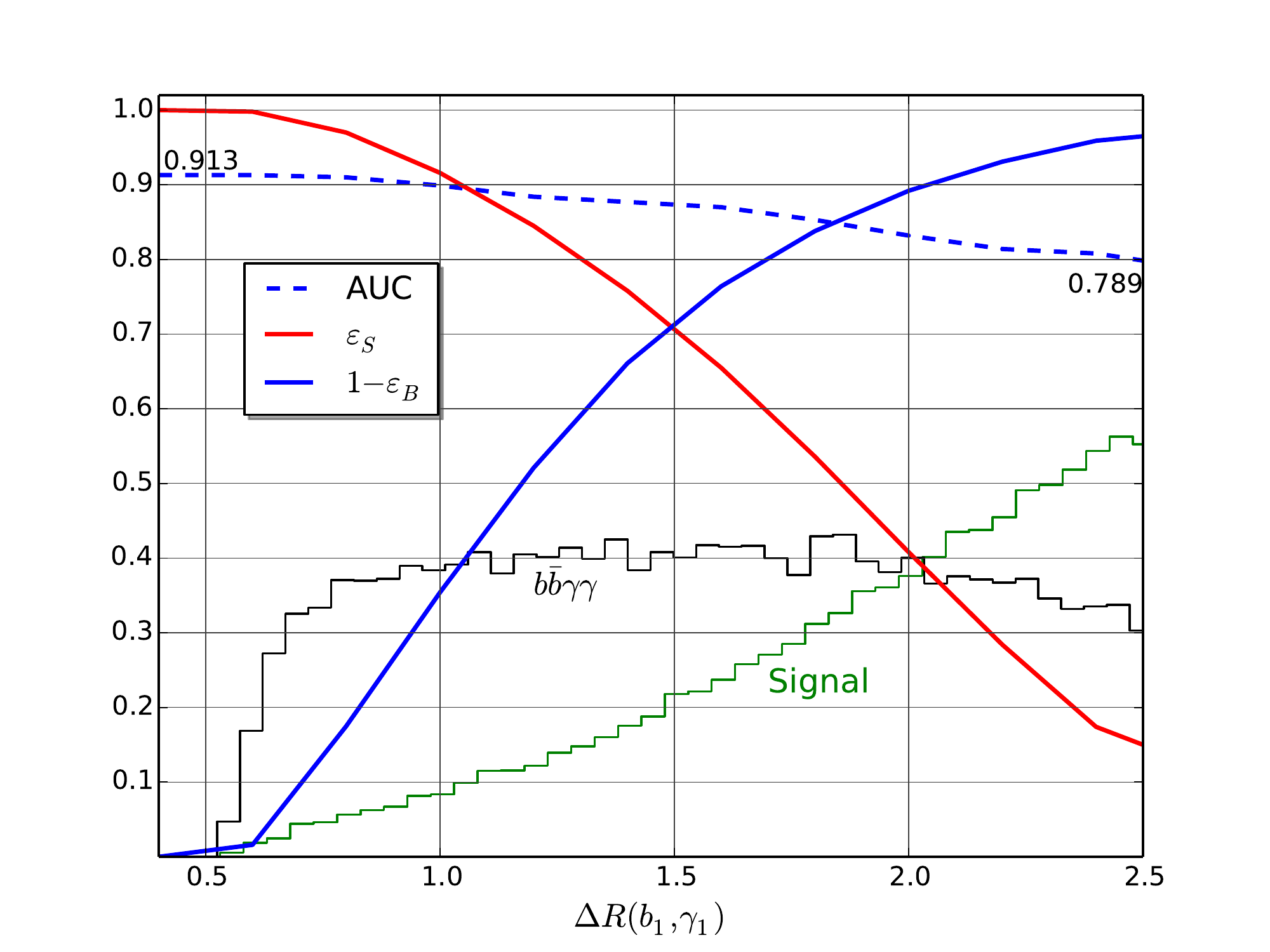}}
\subfigure[\label{fig04}]{\includegraphics[scale=0.35]{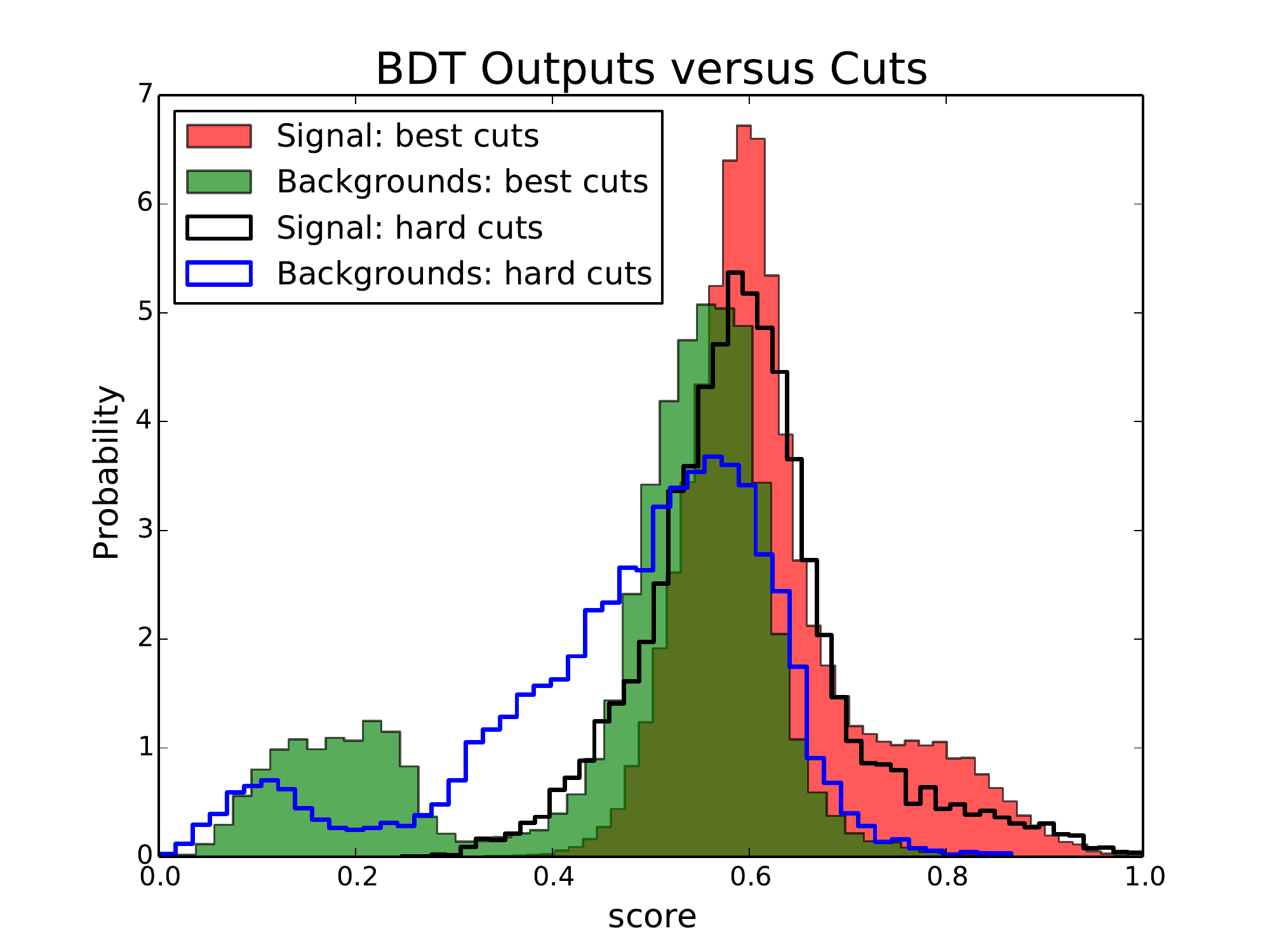}}
\caption{Panels (a) and (b) show the $\Delta R_{b_1\gamma_1}$ distributions of signal and backgrounds requiring the acceptance (default) cuts of eq.~(\ref{cuts:basic}) (Azatov \emph{et. al.}, ref.~\cite{Azatov:2015oxa}, last row of table~\ref{table:1}). In the panel (c) we present the results of the effects of cutting on $\Delta R_{b_1\gamma_1}$ for the BDT performance, see the text for further details. The output scores of the BDT are shown in panel (d) for signal and backgrounds for the optimized set of cuts and a hard set of cuts.}
\label{fig:6}
\end{figure}

We next go on to an investigation of how the cuts affect the discrimination power of the BDT. We fixed the set of cuts as the default cuts of Azatov \emph{et. al.} shown in table~\ref{table:1}, except the $\Delta R_{ij}$ variables. The panels (a) and (b) of figure~\ref{fig:6} show the $\Delta R_{b_1\gamma_1}$ distribution with just the acceptance cuts of eq.~(\ref{cuts:basic}) and after imposing the default cuts of Azatov \emph{et. al.}, respectively. Interestingly, the cuts seem to make the distributions more distinctive in this case; contrary to intuition, therefore, the cuts may help the ML classification in some cases. 

The panel (c) of figure~\ref{fig:6} shows the normalized $\Delta R_{b_1\gamma_1}$ histograms for the signal and the $\bbaa$ continuum background, the signal efficiency(background rejection) is the red(blu) line, and the area under the Receiver-Operator curve (ROC), AUC, is the dashed line. The bigger the AUC, the better the performance of a cut-and-count analysis based on that distribution. 
%
%
To eliminate backgrounds we should demand that an event has a large $\Delta R_{b_1\gamma_1}$. The effect of hardening this cut is that the total background rejection increases and the signal efficiency decreases as expected. For example, requiring $\Delta R_{b_1\gamma_1} > 1.5$, as in the default cuts of Azatov \emph{et. al.} almost exactly rejects 70\% of backgrounds at the same time that it retains 70\% of signal. However, as the cuts get harder the AUC drops from 0.913 to 0.789 as we see from the dashed line. It is common that tiny increments in AUC represent a significant increase in the significance, thus the magnitude of difference in AUC in this case represents a large decrease in the ML performance.

 The BDT scores distributions for signal and backgrounds are shown in panel (d) of figure~\ref{fig:6}. We chose to place harder cuts to make the signal and backgrounds scores distributions more similar. In fact, the hollow histograms of events with hard cuts overlap more noticeably than the scores of the best set of cuts found with the Bayesian optimization. We note, especially, that the green shaded histogram of backgrounds with best cuts presents a more pronounced hill on the left compared to the hollow blue histogram showing some degradation. This isolated left hill is populated mainly by the reducible backgrounds with charged leptons and missing energy. The signal histograms also show marked differences, especially the right hill of best cuts which disappears from the hard cuts of the red hollow histogram.

We next turn to a discussion of the results for the BDT analysis with optimized cuts.

\subsection{Sequential Search for Optimal Cuts and BDT Hyperparameters}
\label{section:sequential}

In this section, we study how best to perform an optimization of the cut analysis and the selection of BDT hyperparameters, in a sequential manner.

The necessity of tuning BDT hyperparameters before optimizing the cuts arises from the need to avoid overfitting and underfitting. This used to be a costly part of a ML analysis. Beside keeping the complexity of the algorithm under control to achieve a good generalization performance, an efficient way to avoid overfitting is to use a large number of training samples whenever possible. For our ML analysis we simulated $\sim 880000$ events as discussed in the previous section. Depending on the cuts, however, the total number of events usually drops to around $100000$--$300000$ events which also turned out to be a sufficient number of samples to keep overfitting under control.

Our first approach was to apply the default cuts of Azatov \emph{et. al.}, and run 500 \HyperOpt trials in the space of the chosen hyperparameters of \xgb in the search for the highest AUC  over 1/3 of the total samples, the other 2/3 were used for  a 5-fold cross validation by randomly splitting the remaining samples in the 2:1 proportion for training and testing the BDT, respectively. The hyperparameters chosen were the \texttt{number of boosted trees}, from 100 to 500, the \texttt{learning rate} from 0.001 to 0.5, the \texttt{maximum depth of the trees}, from 2 to 15 final leaves, and the minimum sum of instance weight needed in a child to continue the splitting process of the tress, \texttt{min\_child\_weight}, from 1 to 6. Once we found the best hyperparameters, we then checked the learning curves of the algorithm, as the classification error and the log-loss, to confirm that it generalizes well from the training to the testing samples.

From this initial tuning we fixed:
\begin{eqnarray} 
\hbox{\texttt{number of boosted trees}} = 200,\; & & \hbox{\texttt{learning rate}} =  0.1\nonumber \\
\hbox{\texttt{maximum depth of the trees}} = 6,\; & & \hbox{\texttt{min\_child\_weight}} = 1. 
\label{param_default}
\end{eqnarray}

Hyperparameters like the \texttt{number of boosted trees}, \texttt{maximum depth of the trees} and the \texttt{min\_child\_weight} are directly related to the complexity of the algorithm by controlling the number, size and configuration of the trees. The \texttt{learning rate}, also known as \texttt{shrinkage} in this context, is a parameter that controls the weight new trees have to further model the data. A large value permits a larger effect from new added trees and might lead to more severe overfitting. There are other parameters which can be eventually used to prevent overfitting and loss of generalization power as explained in refs.~\cite{TChen, Brownlee}, but we found that tuning these parameters was sufficient to achieve a good performance.

In principle, it would also be possible to tune the BDT for each set of cuts. But that would be computationally expensive. As we show going forward, keeping these parameters fixed already leads to very good results in terms of signal significance.

\begin{figure}[!t]
\centering
\includegraphics[scale=0.5]{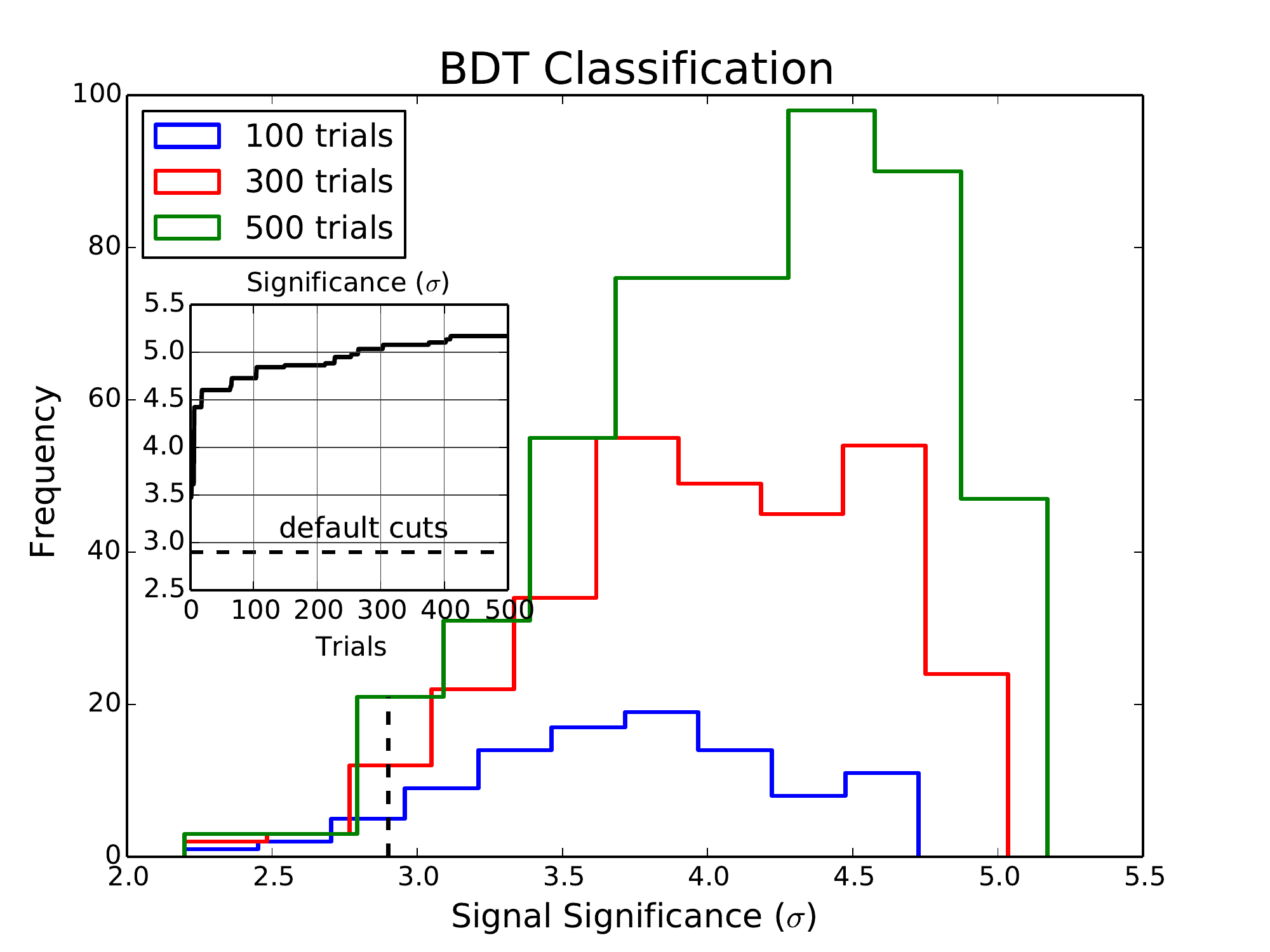}
\caption{The histogram of number of cut strategies producing a given significance interval in a BDT-aided classification analysis. The inset plot shows the significance as a function of the number of \HyperOpt trials. No systematics are assumed, the backgrounds are those of ref.~\cite{Azatov:2015oxa} and the $S/\sqrt{B}$ used to compute the signal significances. The black dashed line represents the results obtained with the default cuts of Azatov \emph{et. al.}, ref.~\cite{Azatov:2015oxa}.}
\label{fig:8}
\end{figure}

In figure~\ref{fig:8} we repeat the analysis presented in figure~\ref{fig:2}, but now after performing the BDT classification. The black dashed line is the maximum signal significance encountered by cutting on the BDT output scores distributions of signal and backgrounds with no systematics using the $S/\sqrt{B}$ metric for the default cuts of Azatov \emph{et. al.} In this case we take only the backgrounds of the ref.~\cite{Azatov:2015oxa} for the comparison.

As in the case of the cut-and-count analysis with no BDT classification, we used \HyperOpt to search for the best cuts in 500 experiments with the same kinematic variables and ranges of table~\ref{table:2}. Again, around 90\% of all cut strategies produced a signal significance larger than the default cuts of Azatov \emph{et. al.} With 200 experiments, the maximum AMS found with the optimized search was 4.9$\sigma$ and an AUC of 0.904, whereas for the default cuts the maximum AMS significance is 2.9$\sigma$ with an AUC of 0.869. The best set of cuts found with 200 experiments was
\begin{eqnarray}
 & & p_T(1) >52\; \hbox{GeV},\; p_T(2) >22\; \hbox{GeV} \nonumber \\
 & & \Delta R_{ij}>0.1,\; \Delta R_{ii}<2.75 \nonumber \\
 & & M_{\bbaa}>340\; \hbox{GeV},\; p_{T_{ii}}>145\; \hbox{GeV},\; M_{b_1\gamma_1}>45\; \hbox{GeV}\nonumber \\
 & & |M_{bb}-m_h|<26\; \hbox{GeV},\; |M_{\gamma\gamma}-m_h|<10\; \hbox{GeV}
\label{cuts1}
\end{eqnarray}

Including the extended backgrounds, again after 200 trials, \HyperOpt found a significance of 4.5$\sigma$, AUC of 0.910, and for the default cuts of Azatov \emph{et. al.}, an AMS of 2.6$\sigma$ and AUC of 0.869. In this case, the Bayesian optimization algorithm found another way into the variables space
\begin{eqnarray}
 & & p_T(1) >52\; \hbox{GeV},\; p_T(2) >20\; \hbox{GeV} \nonumber \\
 & & \Delta R_{ij}>0.65,\; \Delta R_{ii}<3.85 \nonumber \\
 & & M_{\bbaa}>90\; \hbox{GeV},\; p_{T_{ii}}>160\; \hbox{GeV},\; M_{b_1\gamma_1}>125\; \hbox{GeV}\nonumber \\
 & & |M_{bb}-m_h|<24\; \hbox{GeV},\; |M_{\gamma\gamma}-m_h|<12\; \hbox{GeV}
\label{cuts2}
\end{eqnarray}

There is an enormous gain in the significance after using BDT to help classifying signal and background events. However, the $S/\sqrt{B}$ metric overestimates the significance when the number of signal events is not much smaller than the number of background events. In table~\ref{table:4} we show the maximum signal significance by cutting on the BDT scores with extended backgrounds and using the more conservative and best suited significance AMS of eq.~(\ref{sig:ams}). We display in this table the results for all the cut strategies of table~\ref{table:1} plus the best cut strategy found with the Bayesian method.

First, whatever the cut strategy, the BDT classification significantly enhances the signal significance compared to the simple cut-and-count analysis. The larger AMS, however, is once again the one obtained by selecting the cut strategy with the optimized search, reaching $\sim 3.9\sigma$ with 3 ab$^{-1}$ of integrated luminosity. It is interesting to note that the selection cuts found for AMS are different from those of eq.~(\ref{cuts2}) for $S/\sqrt{B}$
\begin{eqnarray}
 & & p_T(1) >92\; \hbox{GeV},\; p_T(2) >20\; \hbox{GeV} \nonumber \\
 & & \Delta R_{ij}>0.2,\; \Delta R_{ii}<2.6 \nonumber \\
 & & M_{\bbaa}>10\; \hbox{GeV},\; p_{T_{ii}}>125\; \hbox{GeV},\; M_{b_1\gamma_1}>70\; \hbox{GeV}\nonumber \\
 & & |M_{bb}-m_h|<30\; \hbox{GeV},\; |M_{\gamma\gamma}-m_h|<9\; \hbox{GeV}
\label{cuts3}
\end{eqnarray}

From the previous results and those of eqs.~(\ref{cuts1}, \ref{cuts2}, \ref{cuts3}) we observe that the Bayesian optimization algorithm learns basically two types of selection criteria to increase the significance: either relaxing the $\Delta R$ and hardening some of the invariant masses and transverse momenta variables, or placing more stringent $\Delta R$ and relaxing invariant mass and transverse momentum cuts. This is perfectly understandable from the physics point of view: events with high $p_T$ particles and large invariant masses are more likely to contain collimated photons and $b$-jets, thus cutting both on $\Delta R$ and invariant masses, for example, would be redundant as also can be seen in panels (a) and (b) of figure~\ref{fig:4}. The job of the optimization algorithm is more a fine tuning of the cuts throughout the variables space.

Another feature of the best cut criteria found so far by the Bayesian approach is the $b$-tagging dependence with the transverse momentum as parametrized in the \Delphes detector simulator. Once the $b$-tagging increases with the bottom quark transverse momentum, selection criteria with at least one high-$p_T$ is likely to provide a better discrimination against important non-$b$ jet backgrounds as $\jjaa$, $\ccaa$ and $\ccaj$.

\begin{table}[t]
\centering
\begin{tabular}{c|c|c}
\hline
Reference & max AMS($\sigma$) with BDT & AUC \\ 
\hline \hline
(A) \cite{Baur:2003gp} & 2.36 & 0.884 \\
\hline
(B) \cite{Baglio:2012np} & 1.96 & 0.885 \\ 
\hline
(C) \cite{Huang:2015tdv} & 2.43 & 0.885 \\
\hline
(D) \cite{Azatov:2015oxa} & 2.65 & 0.870 \\
\hline
ATLAS \cite{ATLAS14} & 2.67 & 0.883 \\
\hline
Our work (with \HyperOpt) & {\bf 3.88} & {\bf 0.901} \\
\hline\hline
\end{tabular}
\caption{Comparison of the performance of the BDT implementation in \xgb trained with samples selected with the cuts various previous works in the literature and with the optimized set of cuts. The second column contains the maximum AMS obtained by cutting on the BDT outputs after a 5-fold cross validation. The last column displays the AUC metric of the BDT for each set of cuts.}
\label{table:4}
\end{table}

\subsection{Joint Search for Best Cuts and BDT Hyperparameters}
\label{section:joint}
In the previous section we carried out a sequential search for cuts and BDT hyperparameters, first adjusting the BDT to perform well on the baseline selection criteria, then, with the hyperparameters fixed, continuing to the search of best cuts.

In this section, we will investigate whether a joint search for all the parameters of the phenomenological analysis can also yield good results. The relevant parameters that need to be adjusted together are both the cut thresholds and the BDT hyperparameters. This represents a more thorough approach to the problem of getting the best performance possible using a ML algorithm.

For this global search we used \HyperOpt with the parameters space of the table~\ref{table:5}. All the prior distributions were assumed to be uniform in the range indicated in the right column. As in the previous analysis, the objective function to be minimized was $-AMS$. All the BDT results were obtained from a 5-fold cross validation by randomly splitting training and testing samples at the proportion of 2/3 and 1/3 of the total sample, respectively. As the parameter space is larger now, we allowed for 300 trials. 

\begin{figure}[!t]
\centering
\includegraphics[scale=0.5]{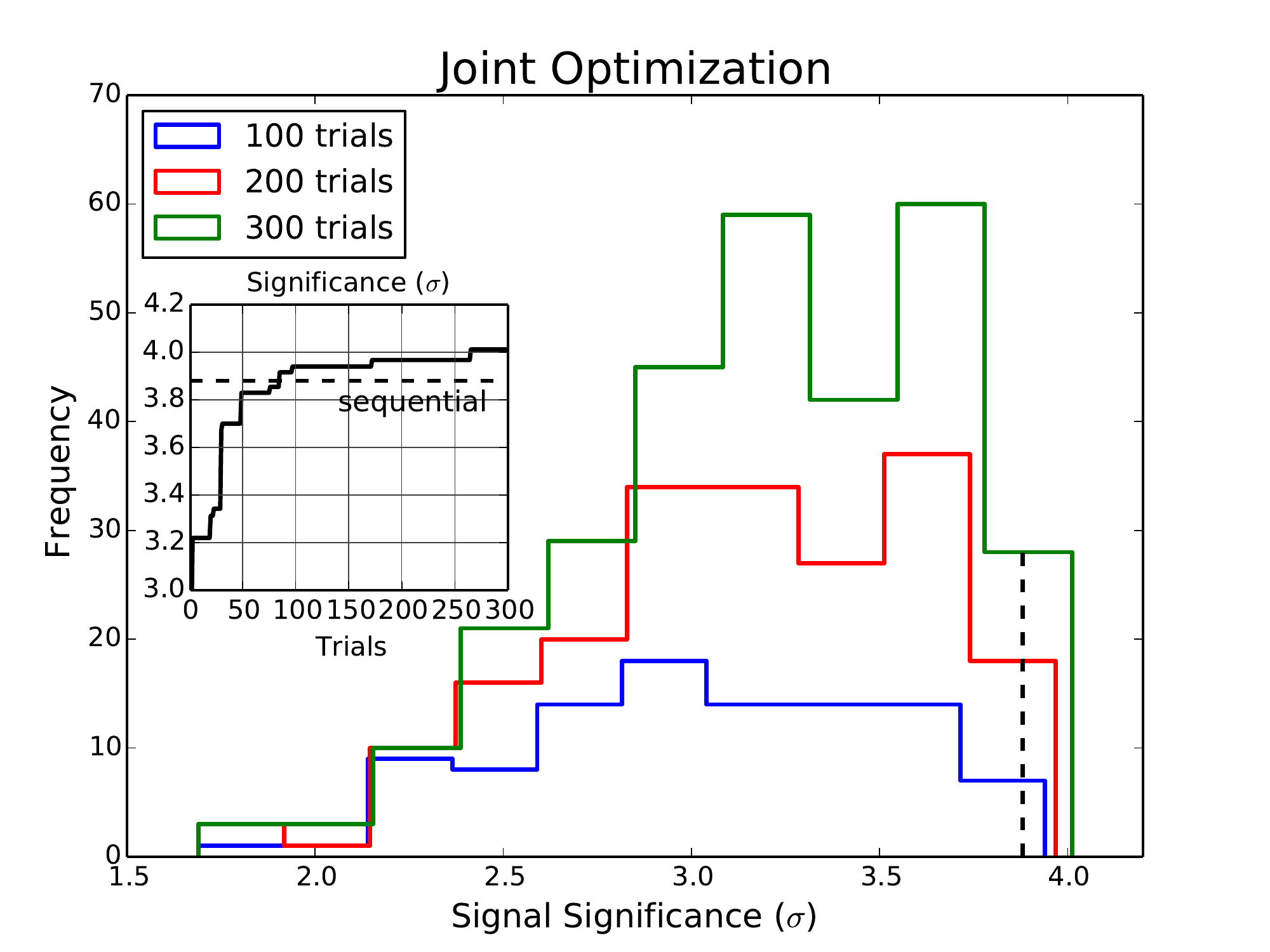}
\caption{The histogram of number of cut strategies producing a given significance interval with BDT adjusted in a joint optimization of cuts and hyperparameters. The inset plot shows the significance as a function of the number of \HyperOpt trials. No systematics are assumed, the backgrounds are those of ref.~\cite{Azatov:2015oxa} and the $S/\sqrt{B}$ used to compute the signal significances. The black dashed line represents the results obtained with the default cuts of Azatov \emph{et. al.}, ref.~\cite{Azatov:2015oxa}.}
\label{fig:9}
\end{figure}

As in the previous sections, we plot, in figure~\ref{fig:9}, histograms of the number of cut strategies for a given significance interval for the joint search. The black dashed line now represents the maximum significance of the sequential search of the previous section. In contrast to the other cases, the global search found just a few better cut strategies, but the important fact is that it actually found a better strategy than the sequential search, showing that a joint search is not only possible but also beneficial to the AMS maximization.

The maximum AMS is 4.0$\sigma$ for an AUC of 0.904, against 3.9$\sigma$ and AUC of 0.901 of the sequential parameters search. The parameters of the joint search are the following
\begin{eqnarray}
 & & p_T(1) >72\; \hbox{GeV},\; p_T(2) >20\; \hbox{GeV} \nonumber \\
 & & \Delta R_{ij}>0.15,\; \Delta R_{ii}<3.6 \nonumber \\
 & & M_{\bbaa}>370\; \hbox{GeV},\; p_{T_{ii}}>145\; \hbox{GeV},\; M_{b_1\gamma_1}>100\; \hbox{GeV}\nonumber \\
 & & |M_{bb}-m_h|<27\; \hbox{GeV},\; |M_{\gamma\gamma}-m_h|<11\; \hbox{GeV}\nonumber \\
 & & \hbox{\texttt{number of trees}}=157 \nonumber \\
 & & \hbox{\texttt{learning rate}}=0.101 \nonumber \\
 & & \hbox{\texttt{maximum tree depth}}=14 \nonumber \\
 & & \hbox{\texttt{min\_child\_weight}}=5
\label{cuts4}
\end{eqnarray}
\begin{table}[t]
\centering
\begin{tabular}{c|c}
\hline
Kinematic variable/BDT Hyperparameter & Variation range in \HyperOpt\\
\hline\hline
$\Delta R_{ii} <$ & $(1.4,4,0.05)$ \\
\hline
$\Delta R_{ij} >$ & $(0,2,0.05)$ \\ 
\hline 
$p_T(1) >$ & $(30,100,1)$ GeV \\
\hline 
$p_T(2) >$ & $(20,70,1)$ GeV \\
\hline 
$p_{T_{ii}} >$ & $(0,200,5)$ GeV \\
\hline 
$M_{\bbaa} >$ & $(0,400,5)$ GeV \\
\hline 
$M_{b_1\gamma_1} >$ & $(0,200,5)$ GeV \\
\hline 
$|M_{\gamma\gamma}-m_h| <$ & $(5,15,1)$ GeV \\
\hline 
$|M_{bb}-m_h| <$ & $(10,30,1)$ GeV \\
\hline
\texttt{number of trees} & $(150,250,1)$ \\
\hline
\texttt{learning rate} & $(0.001,0.5,0.001)$ \\
\hline
\texttt{maxixum tree depth} & $(2,20,1)$ \\
\hline
\texttt{min\_child\_weight} & $(1,6,1)$ \\
\hline\hline
\end{tabular}
\caption{The kinematic variables used for cuts and BDT hyperparameters and their allowed variation ranges in \HyperOpt for the joint optimization. The prior distributions for all these variables are set to uniform distributions over the ranges shown in the table within the steps shown as the last entry of each vector. In a grid search, the number of evaluation points would be approximately $8.9\times 10^{20}$.}
\label{table:5}
\end{table}

The joint search was able to find a more regularized set of BDT hyperparameters to avoid overfitting. This is why the \texttt{number of trees} is smaller and \texttt{min\_child\_weight} bigger than those of the sequential search. This is a very welcome result - with harder cuts than those of the default cuts of Azatov \emph{et. al.}, \HyperOpt learned how to control the loss in AMS that would be caused by a more dangerous overfitted BDT due a smaller size sample for training and testing. Moreover, it was able to tune the parameters to perform slightly better than the sequential search. Finally, we note that the optimized cuts are of the type that feature harder invariant masses and transverse momenta and relaxed $\Delta R$ cuts.

As a final investigation, we present in the next section a multivariate statistical analysis based on the BDT output scores and the inclusion of systematic uncertainties for our final more realistic prospects of discovering the double Higgs production at the LHC.

\section{Final Results: Further Discrimination with Multivariate Analysis of BDT Outputs}
\label{section:mva}

In the previous sections, we  employed a ML algorithm to boost our classification accuracy of signal and background events, relying exclusively on cut-and-count analysis and posterior calculation of the significance with an approximated median significance formula. 

In this section, we will attempt to improve the signal significance by focusing on the statistical side of the analysis encouraged by the results of ref.~\cite{Kling:2016lay}, around $4\sigma$ for 3 ab$^{-1}$ with MVA based on kinematic variables but with no systematics included. This will be done by estimating the log-likelihood ratio statistics from the output scores of the BDT algorithm provided by \xgb. This is a well known and established procedure used by the LHC Collaborations for a long time, but only recently more rigorously justified~\cite{Cranmer:2016swd}. 

 We calculate the log-likelihood ratio of the binned BDT output scores for signal $s_i$ and backgrounds $b_i,\; i=1,\cdots, N_{bins}$, after cuts, shown in the panel (d) of figure~\ref{fig:6}, according to~\cite{Cowan:2013pha}
\begin{equation}
\Lambda = \sum_{i=1}^{N_{bins}} \left[-s_i+d_i\ln\left(1+\frac{s_i}{b_i}\right)\right]
\label{llr}
\end{equation}

We assume that the simulated data $d_i$ follows either the null hypothesis with no signal, $d_i\sim Pois(x_{BDT_i}|B)$, to compute $\Lambda_B$, or the alternative hypothesis where $d_i\sim Pois(x_{BDT_i}|S+B)$ to compute $\Lambda_{S+B}$. The estimation of the non-parametric statistical distributions of $\Lambda_B$ and $\Lambda_{S+B}$, $P(\Lambda|B)$ and $P(\Lambda|S+B)$, respectively, is done with a large number of pseudoexperiments with new statistically varied BDT output distributions assuming that the number of events in each bin is drawn from a Poisson distribution, $Pois(x|\mu)$, of mean $\mu$. From these distributions the $p$-value of the background hypothesis is calculated
\begin{equation}
p_B=\int_{\Lambda_{S+B}}^{+\infty} P(\Lambda|B)d\Lambda
\label{p-value}
\end{equation}
and the statistical significance is computed as $\Phi^{-1}(1-p_B)$, where $\Phi$ is the cumulative distribution function of the standard Gaussian with zero mean and unit variance.

According to the Neyman-Pearson lemma~\cite{NP}, the likelihood ratio is the most powerful test statistic to discriminate a signal hypothesis for a fixed significance level of the background hypothesis (a fixed background efficiency) in the absence of systematic uncertainties. 

In this work, in order to estimate $P(\Lambda|B)$ and $P(\Lambda|S+B)$, we performed 40000 pseudoexperiments from the binned BDT output scores. As in the previous sections, we used \HyperOpt to search for the cut strategy with the biggest significance after training the BDTs and computing the AMS as described above. The BDT hyperparameters were fixed as in eq.~(\ref{param_default}), so Bayesian search was applied in the sequential way.

\begin{figure}[!t]
\centering
\includegraphics[scale=0.5]{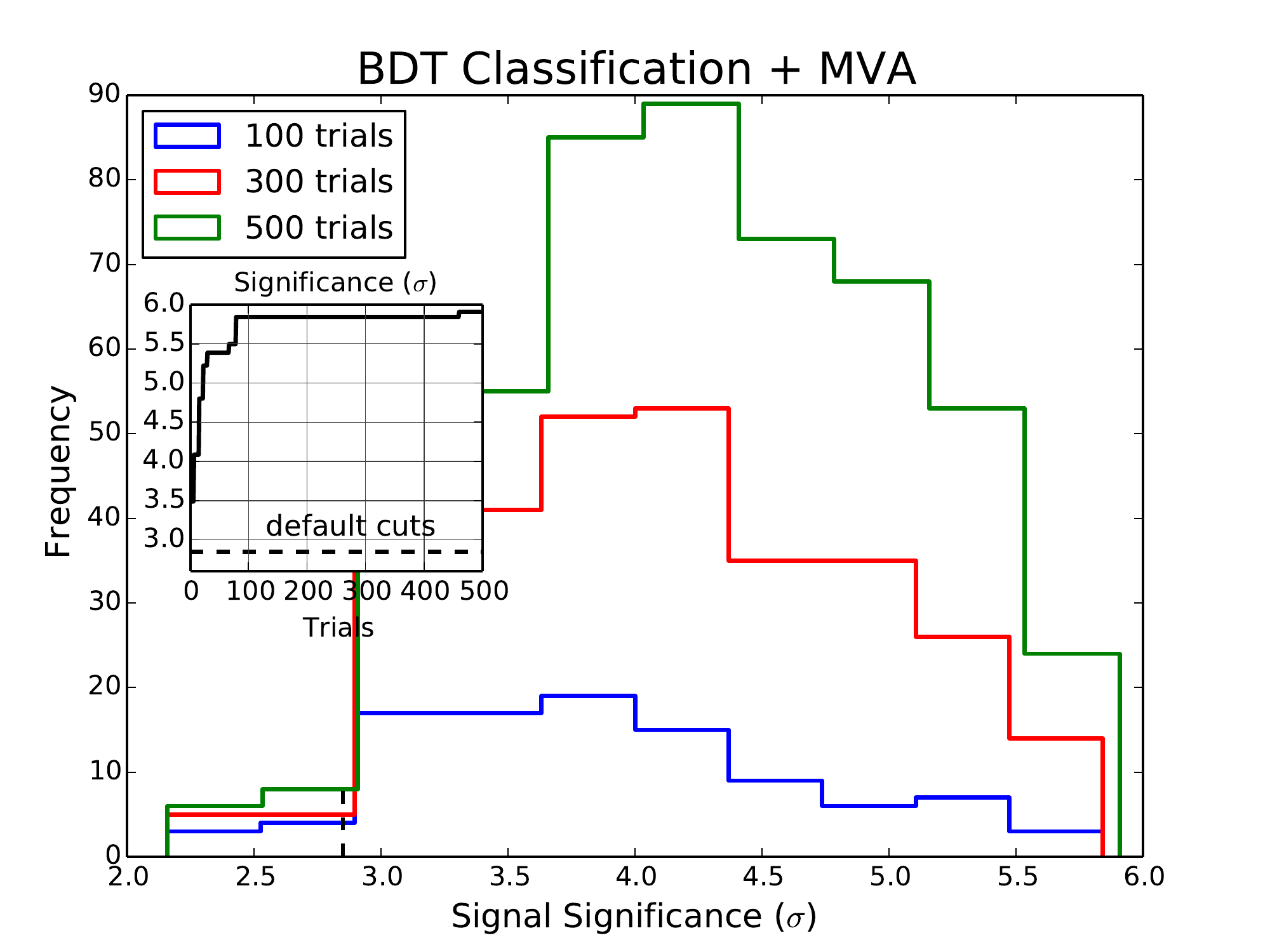}
\caption{The histogram of number of cut strategies producing a given significance interval in MVA. The inset plot shows the significance as a function of the number of \HyperOpt trials. No systematics are assumed, the backgrounds are those of ref.~\cite{Azatov:2015oxa} and the $S/\sqrt{B}$ used to compute the signal significances. The black dashed line represents the results obtained with the default cuts of Azatov \emph{et. al.}, ref.~\cite{Azatov:2015oxa}. The optimization was of the sequential type.}
\label{fig:10}
\end{figure}

The histogram of cut strategies as a function of the significance of figure~\ref{fig:10}, as in the other cases, shows that more than 90\% of all cut selections found by \HyperOpt lead to a better MVA performance than that of the default cuts of Azatov \emph{et. al.}, which are definitely not suited to MVA. Also, similarly to other cases studied previously, the maximum significance is found rather early in the searching, with 100 experiments, as shown in the inset plot of figure~\ref{fig:10}. A very high AMS is already obtained at that stage, and it is the best strategy up to almost the 500th experiment which improves it very slightly.

Our final analysis and results take into account systematic uncertainties of 10\% and 20\% and are shown in the table~\ref{table:6}. The systematic uncertainties are incorporated in MVA in mixed frequentist-Bayesian method, by marginalizing over the background rate in eq.~(\ref{p-value}) assuming that the systematic errors are Gaussian. All the backgrounds are taken into account, including $\ccaa$, $\bbaj$ and $\ccaj$, the significance was calculated with the AMS formula~(\ref{sig:ams}), and the integrated luminosity corresponds to 3~ab$^{-1}$.

 With a very low level of systematics, the techniques proposed here with the selection criteria optimization may be able to confirm the production of a pair of SM Higgs bosons with 5$\sigma$. Within a more realistic projection of the level of systematics, around 10\%, the optimization of cuts to train boosted decision trees combined with a multivariate analysis delivers a respectable significance of 4.6$\sigma$. This is the largest significance achieved so far in the $\bbaa$ channel with realistic assumptions concerning backgrounds and systematic uncertainties at the 14 TeV LHC. Even assuming large systematics of 20\%, our analysis predicts a 3.6$\sigma$ significance, which represents at least a strong evidence in favor of double SM Higgs production.

Relying just on BDT classification with optimized cuts, for systematics below 20\%, a robust evidence for double Higgs production is possible according to table~\ref{table:6}. 

Compared to the default cuts of ref.~\cite{Azatov:2015oxa}, which we took in this work as our baseline results, the cuts found from the Bayesian optimization are able to enhance the significance by 30\%--50\% with little computational efforts and speed. The results for the default cuts of Azatov \emph{et. al.} are shown between brackets in the second column of table~\ref{table:6} for comparison.

Finally, we elect from all the results presented, those of the second row of table~\ref{table:6} as the most representative of our findings, again stressing that these results take into account realistic backgrounds, the level of systematic uncertainties expected for this channel, and also better suited significance metrics for the number of signal and background events expected at the LHC with these selection criteria.


\section{Conclusions and Prospects}
\label{section:conclusions}
\begin{table}[t]
\centering
\begin{tabular}{c|c|c|c}
\hline
systematics (\%) & Cut-and-count & BDT & MVA \\
\hline\hline 
0 & 2.34[1.76] & 3.88 & 5.05 \\
\hline
10 & {\bf 1.93}[1.43] & {\bf 3.57} & {\bf 4.64} \\
\hline
20 & 1.51[1.0] & 3.10 & 3.60 \\
\hline\hline
\end{tabular}
\caption{Signal significances for cut-and-count, BDT and MVA are shown in the second, third and fourth column, respectively, for 0, 10 and 20\% systematics. We took all backgrounds into account for the computation of the AMS with optimized cuts and an integrated luminosity of 3 ab$^{-1}$ at the 14 TeV LHC. The bold-face numbers represent the significances expected with the level of systematics anticipated by the experimental collaborations in refs.~\cite{ATLAS14, ATLAS17, CMS}. The numbers inside brackets are the significances computed with the default cuts of Azatov \emph{et. al.}, ref.~\cite{Azatov:2015oxa}, which we took as baseline results.}
\label{table:6}
\end{table}

In this paper, we explored double Higgs production via gluon fusion at the LHC. Our analysis builds significantly on previous studies in that we used tools from the ML literature to discriminate signal and background events. We also incorporated  background contributions coming from light flavor jets or $c$-jets being misidentified as $b$-jets and electrons or jets being misidentified as photons.

First we used Bayesian optimization, implemented in \HyperOpt , to select cuts on kinematic variables, obtaining a $30-50$ \% increase in the significance metric $S/\sqrt{B}$ compared to current results in the literature. Then, we used BDTs implemented in \xgb to further discriminate signal and background events. At this stage, we showed that a joint optimization of kinematic cuts and BDT hyperparameters results in an appreciable improvement in performance. Finally, we turned to the statistical side of the analysis by estimating the log-likelihood ratio statistics from the output scores of the BDT algorithm provided by \xgb . The final results of our paper are presented in table~\ref{table:6}. We find that assuming a very low level of systematics, the techniques proposed here will be able to confirm the production of a pair of SM Higgs bosons at 5$\sigma$ level. Assuming a more realistic projection of the level of systematics, around 10\%, the optimization of cuts to train BDTs combined with a multivariate analysis delivers a respectable significance of 4.6$\sigma$. This is the largest significance achieved so far in the $\bbaa$ channel with realistic assumptions concerning backgrounds and systematic uncertainties at the 14 TeV LHC. Even assuming large systematics of 20\%, our analysis predicts a 3.6$\sigma$ significance, which represents at least strong evidence in favor of double SM Higgs production.

We pause for a moment to recapitulate the reasons behind the larger significances obtained in this paper, compared to previous studies. What makes the significances larger is precisely the better discrimination between the signal and background classes achieved by the machine learning algorithms as they find more profound correlations among the kinematic features and those classes. These correlations cannot be fully explored in simple/manual rectangular cut-and-count analyses. There is a tradeoff between the efficiency of the cuts and the ML performance which is usually neglected in phenomenological works where these tools are employed. The reasoning is simple: cutting harder cleans up more backgrounds but weakens the correlations between the kinematic variables and the event classes, thereby decreasing the ML performance. On the other hand, relaxing the cuts makes the correlations stronger helping to boost ML but the discrimination power gained might not be enough to get a good significance with a large number of surviving background events. Finding the optimal performance from this competition is the core of the method present in the paper.

We now turn to some future prospects. One immediate future goal is to study the prospects of measuring deviations of $\lambda_3$ from the SM prediction at the high-luminosity LHC using our work on double Higgs processes in the $\bbaa$ channel \cite{wip}. In this context, it would also be interesting to pursue the ensuing implications for the electroweak phase transition within an effective potential framework. Another set of goals is to extend our work to other final states like $\bbtautau$, $\bbWW$, and $\bbbb$, as well as other production channels. 

There are also several directions one can pursue that are not necessarily related to studies of the Higgs sector. The Bayesian optimization approach to the cut selection presented in this work can be used in other phenomenological studies. For example, it would be very interesting to use our methods to re-evaluate the discovery prospects for compressed supersymmetric searches or dark matter ~\cite{Dutta:2013gga, Delannoy:2013ata,Dutta:2014jda}. The Bayesian optimization can also be used to design a cut selection that helps to overcome the effect of various types of systematic uncertainties which affect the shape of the distributions and the normalization of the cross sections.

The measurement of particles masses, couplings and quantum numbers like spin and CP also depend strongly on the kinematic selection criteria. This is another target for optimization using \HyperOpt. As we showed in this work, a multivariate analysis used for hypothesis tests can be greatly enhanced with a careful set of cuts aimed to keep strong correlations but eliminating as much backgrounds as possible. Some other discrimination techniques which suffer with hard cuts, such as the calculation of asymmetries~\cite{Barr:2005dz, Alves:2007xt, Alves:2006df, Smillie:2005ar, Athanasiou:2006ef}, are also worth investigating using our methods.

While we performed a discovery analysis in this work, an optimized set of cuts, with or without further classification with the help of ML tools, can also be employed to obtain stringent limits in exclusion studies.

It is certain that cut optimization will be able to improve the performance of other classifiers such as neural networks and naive Bayes-inspired algorithms which are commonly explored in phenomenological studies, although it is difficult to estimate the extent. One might anticipate that the cut selection which optimizes a given classifier should not correspond to the selection that improves another. The joint optimization presented here is also a potential target of further investigation as the cut optimization can be performed at the same time of hyperparameters tuning of classifiers as decision trees and neural networks. These are directions that can also be pursued in the future.

\section*{Acknowledgments}

A. Alves would like to thank Funda\c{c}\~ao de Amparo \`a Pesquisa de Estado de S\~ao Paulo (FAPESP), process 2013/22079-8, and Conselho Nacional de Desenvolvimento Cient\'ifico e Tecnol\'ogico, CNPq, grant 307098/2014-1. K. Sinha would like to thank Minho Son for a very illuminating discussion, and the IBS Center for Theoretical Physics of the Universe, Daejeon, South Korea for hospitality when this work was in progress. He would also like to thank Paul Padley, Ankit Patel, and Jamal Rorie for helpful discussions. TG is supported by the United State Department of Energy Grant Number de-sc 0016013.

\appendix

\section{Statistical significance metrics}
\label{app:1}
A comparative study of various statistical significance metrics can be found in~\cite{Cousins}. In that work, the problem of incorporating systematic uncertainties in the background normalization for a Poisson process is addressed
and it is found that three most widely used significance metrics perform similarly in many situations concerning the relative number of signal and background events and the level of systematics.

The three significance methods are:
\begin{itemize}
\item[(1)] The naive and most simple way to incorporate systematic uncertainties in the calculation of the significances for $S$ signal events and $B$ background events in a Poisson process for a given integrated luminosity
\begin{equation}
\frac{S}{\sqrt{B+(\varepsilon_B B)^2}}
\label{sig:naive}
\end{equation}

In all cases, we assume that the systematic uncertainty in the total background normalization is proportional to the number of background events, $\varepsilon_{B} B$. This is simple and fast, but it somewhat overestimates the discovery reach with or without systematics.

\item[(2)] The Bayesian-frequentist hybrid recipe to the estimation of the systematics impact on the significance. Assuming that systematic errors are normally distributed we marginalize over the systematic errors to obtain the $p$-value 
\begin{equation}
p_B=\sum_{k=S+B}^{+\infty} \int_{-\infty}^{+\infty} \frac{e^{-B(1+z\varepsilon_B)}}{k!}\left[B(1+z\varepsilon_B)\right]^k\times \frac{e^{-\frac{z^2}{2}}}{\sqrt{2\pi}}\, dz
\label{sig:mva}
\end{equation}
and the significance is computed as $Z=\Phi^{-1}(1-p_B)$, where $\Phi(z)$ is the cumulative distribution function of the standard normal distribution.

This is method of incorporating systematics into the significance was employed in section~\ref{section:mva} for the MVA analysis and it is computationally more demanding.

\item[(3)] The Profile Likelihood method originally proposed in~\cite{LiMa} in astrophysical searches with subsidiary measurements of the background adapted to a high energy experiment where the systematics is a fraction of background events, $\varepsilon_B B$
\begin{equation}
AMS=\left\{\begin{array}{l}
\sqrt{2}\left\{(S+B)\ln\left[\left(1+\frac{1}{B\varepsilon_B^2}\right)\frac{S+B}{S+B+1/\varepsilon_B^2}\right]
+\frac{1}{\varepsilon_B^2}\ln\left[\frac{B+1/\varepsilon_B^2}{S+B+1/\varepsilon_B^2}\right]\right\}^{\frac{1}{2}},\; \varepsilon > 0\\
\sqrt{2}\left[-S+(S+B)\ln\left(1+\frac{S}{B}\right)\right]^{\frac{1}{2}},\; \varepsilon = 0 
\end{array}\right.
\label{sig:ams}
\end{equation}

Among the three metrics this is the most conservative and reliable, and it is as simple and fast to compute as the naive metrics of eq.~(\ref{sig:naive}). Moreover, its performance is very close to the consistent frequentist approach for tests of the ratio of Poisson means implemented in \texttt{ROOT}~\cite{root}, for example. 
\end{itemize}

A comparison of these three methods are also investigated in ref.~\cite{Alves:2015dya} in the context of the search for dark matter production in the mono-$Z$ channel confirming all the features anticipated in ref.~\cite{Cousins}. In the case of double Higgs production, we also checked that the naive formula of eq.~(\ref{sig:naive}) always provide larger significances with or without systematics compared to the other metrics for the same number of signal and background events. 

\section{Python code of the optimization method using \HyperOpt}
\label{app:2}

We show a snippet of the code used to optimize the cut strategies right below. 
\begin{lstlisting}[language=Python, caption=Python snippet of the optimization code.]
# loading packages
import numpy as np
from hyperopt import hp, fmin, tpe, STATUS_OK, Trials 
from functools import partial
# loading data
data = np.genfromtxt('data/data.csv', delimiter=',')
n_data, ncol = data.shape
print data.shape
ncol=ncol-1
# raw data
X_raw = data[:,1:ncol] #vector of features, 27 for HH production
y_raw = data[:,0]      #labels
weights=data[:,ncol]   #events weights
print ('finish loading '+str(n_data)+' samples from csv file')
# evaluation parameters
nevals=200
# building the selector function
def selector(y):
    aux_min=int(min(y))
    aux_max=int(max(y))
    sel=[[] for i in range(int(aux_max))]
    for i in range(int(aux_max)):
        sel[i] = np.array([y[k] == float(i+1) for k in range(len(y))])
    return sel
# AMS metrics
def ams(s,b,sys):
    breg=0.0
    #return s/np.sqrt(b+breg+(sys*(b+breg))**2)
    #return np.sqrt(2.0)*np.sqrt( (s+b+breg)*np.log(1.0+s/(b+breg))-s )
    if b==0. and sys!=0.:
        aux=np.sqrt(2*s*(1.+1./sys))
    else:
        b=b+breg
        if sys==0.:
            aux=np.sqrt(2.0)*np.sqrt(-s+(s+b)*np.log(1.+s/b))
        else:
            aux=np.sqrt(2.0)*np.sqrt((s+b)*np.log((1.+1./(b*sys**2))*(s+b)/(s+b+1./sys**2))+(1./sys)**2*np.log((1.+b*sys**2)*(1/sys**2)/(s+b+1./sys**2)))
    return aux      
# computing the number of signal, backgrounds events for a given selection
def Nevents(w,sel):
    nev=len(y)
    # number of events of each class
    nevS  = np.sum(np.array([w[sel[0]]])) #signal
    nevB1 = np.sum(np.array([w[sel[1]]])) #background 1
    nevB2 = np.sum(np.array([w[sel[2]]])) #backgriund 2
    nevB3 = np.sum(np.array([w[sel[3]]])) #background 3
    nevB  = nevB1+nevB2+nevB3
    events= np.array([nevS,nevB1,nevB2,nevB3])
    return events  
#############################
# Passcuts Boolean function #
#############################
# variables contained in the vector of features
vars={'pT1':1, 'pT2':2, 'Mii':3, 'Mij':4, 'Rij':5}
# defining cut variables
mh=125.0
vd1=data[:,vars['pT1']]
vd2=data[:,vars['pT2']]
vd3=data[:,vars['Mii']]
vd4=data[:,vars['Mij']]
vd5=data[:,vars['Rij']]
vd6=abs(vd4-vd3+mh*np.ones(n_data))
# cuts function
def passcuts(cut,a):
    if a[0]>=cut[0] and a[1]>=cut[1] and abs(a[2]-mh)<=cut[2] \
       and a[3]>=cut[3] and a[4]<=cut[4] and a[5]>=cut[5]: 
        aux=True
    else:
        aux=False
    return aux
###############################
# CUT-AND-COUNT: TPE/HyperOpt #
###############################
best_cc=[[] for i in range(22)]
best_cut=[[] for i in range(22)]
def objective(cuts):
    cut=np.array([cuts['pT1_cut'],cuts['pT2_cut'],cuts['Wii_cut'], \
                  cuts['Mij_cut'],cuts['Rij_cut'],cuts['Mxx_cut']])
    data_cut=np.array([data[i] for i in range(n_data) if \
                       passcuts(cut,[vd1[i],vd2[i],vd3[i],\
                                     vd4[i],vd5[i],vd6[i]])])
    if len(data_cut)!=0:    
        y_cut = data_cut[:,0]
        n_cut = len(y_cut)
        w_cut = data_cut[:,ncol]
        sel_cut = selector(y_cut)
    # number of events of each class
        nevS, nevB1, nevB2, nevB3 = Nevents(w_cut,sel_cut)
        nevB  = nevB1+nevB2+nevB3
        loss=-ams(nevS,nevB,sys)
        print -loss
    else:
        print 'no events passed cuts'
        loss=0.
    return{'loss':loss, 'status': STATUS_OK}
# Cuts dictionary
cuts={
        'pT1_cut': hp.quniform("pT1_cut", 30., 100., 1.),   #70
        'pT2_cut': hp.quniform("pT2_cut", 20., 60., 1.),    #30
        'Wii_cut': hp.quniform("Wii_cut", 5., 15., 1.),     #10
        'Mij_cut': hp.quniform("Mij_cut", 100., 300., 10.), #20 
        'Rij_cut': hp.quniform("Rij_cut", 0.4, 1.4, 0.1),   #10
        'Mxx_cut': hp.quniform("Mxx_cut", 100., 200., 10.)  #20 
    }
print '-----HyperOpt SEARCH: '+str(nevals)+' experiments ------'
for j in range(0,25,5):
    sys=j*0.01
    print 'systematics = '+str(j)+'%'
    trials = Trials()
    best = fmin(fn=objective,
                space=cuts,
                algo=partial(tpe.suggest, n_startup_jobs=10)#rand. for random search
                max_evals=nevals,
                trials=trials)
    print 'best:'
    print best
    best_cut[j]=best
    # best ams calculation
    cut=np.array([best['pT1_cut'],best['pT2_cut'],best['Wii_cut'], \
                  best['Mij_cut'],best['Rij_cut'],best['Mxx_cut']])
    data_best=np.array([data[i] for i in range(n_data) if \
                        passcuts(cut,[vd1[i],vd2[i],vd3[i], \
                                      vd4[i],vd5[i],vd6[i]])])
    y_best = data_best[:,0]
    n_best=len(y_best)
    w_best = data_best[:,ncol]
    # number of events of each class
    nevS, nevB1, nevB2, nevB3 = Nevents(w_best,y_best)
    nevB  = nevB1+nevB2+nevB3
    best_cc[j]=ams(nevS,nevB,sys)        
    print 'sys, AMS, S/B =', sys, ams(nevS,nevB,sys), nevS/nevB
\end{lstlisting}

This code illustrates the basic steps to optimize a cut strategy with a single signal class and three different background classes as an example. It cannot be immediately used, but should be adapted to the reader analysis.

First, we load the basic Python packages \texttt{NumPy} and \HyperOpt and also load the data from lines 1 to 14. If the data size is too big it might be necessary to load it in batches. In line 16 we set the number of \HyperOpt trials.

Signal and background samples need to be identified in several steps of the computation, we then create an event selector with the event labels as input in line 18. Significance metrics discussed in the previous appendix can be chosen in the definition of the \texttt{ams} function at line 26, $s(b)$ is the number of signal(backgrounds) events and \texttt{sys} the systematics level in the background rate $\varepsilon_B$.

In the line 40 we define a function that returns the number of signal and background events given a selector vector.

From lines 50 to 70 we build a Boolean function which returns \texttt{True} if an event pass the cuts, otherwise it returns \texttt{False}. This function is inspired in the Fortran routine \texttt{PASSCUTS} found in the \texttt{MadAnalysis} package for \texttt{MadGraph}~\cite{MG5}. This function needs to be adjusted by the user according to his/her selection criteria.
In this example we put cuts on all the features of the event but this is not mandatory, of course. Instead, we construct in line 62 another cut variable which does not compound the features vector.

Now comes the part of the code where we actually perform the optimization. At the line 76 we define our \texttt{objective} function which is going to be minimized by \HyperOpt, its input is the cut dictionary placed at line 77. In this case, we are interested in maximize the \texttt{ams} function, that is, minimize \texttt{-ams}. Note that prior to the computation of \texttt{ams} we select those events which pass the cuts designed in \texttt{passcuts}. The labels and weights of these selected events are denoted by \texttt{y\_cut} and \texttt{w\_cut}, respectively. With \texttt{y\_cut} we set the events selector at line 86 and then call the \texttt{Nevents} function to calculate the number of signal and background events after cuts, these numbers feed the \texttt{-ams} function at line 90.

In the line 97 we set the a Python dictionary for the cut thresholds to be chosen by \HyperOpt with the corresponding priors, in this case, all the priors were chosen to be uniform distributions. In ref.~\cite{Bergstra2} the user can find all the options to set the functioning of the program.

At line 106 we start a loop in the systematics level \texttt{sys} from 0 to 20\%, from 5 to 5\%. The TPE search is called in 110 in order to find the best cuts (with a warm-up phase of 10 trials) which return the larger AMS within \texttt{nevals} trials. From line 118 until the line 131 we calculate and print the results of the optimization for a given systematics. If one wants to perform a random search instead of using TPE, the line 112 should be modified to \texttt{algo = rand.suggest}.

Note that the quantile $\gamma$ discussed in section~\ref{section:smbo} is, in principle, an adjustable parameter, but as far as we know there is no option to change it in \HyperOpt. In ref.~\cite{Bergstra1}, however, the authors keep this parameter at 0.15 for their studies.


\end{document}